\documentclass[twocolumn]{aastex631}
\makeatletter
\renewcommand{\frontmatter@title@above}{}
\makeatother
\usepackage{amsmath}
\usepackage{amssymb}
\usepackage{soul}
\usepackage{todonotes}
\usepackage{multirow}
\usepackage{color, colortbl}
\usepackage{tabularx}
\definecolor{Gray}{gray}{0.95}
\usepackage{graphicx}
\usepackage{booktabs}
\usepackage{rotating}
\usepackage{comment}
\usepackage{subfigure}
\usepackage{tabularx}
\usepackage{hyperref}
\usepackage{float}
\usepackage{needspace}
\usepackage[normalem]{ulem}

\usepackage[normalem]{ulem}

\newcommand{\ba}{\begin{array}}
\newcommand{\ea}{\end{array}}

\begin{document}

\title{The Impact of Galaxy Catalogue Completeness with GLADE+ on Dark Siren Cosmology: A Line-of-Sight Prior Analysis of GWTC-3 Events}

\author[0009-0009-5450-6318]{Juan Esteban Rivera}\affiliation{Instituto de Fisica, Universidad de Antioquia, A.A.1226, Medellin, Colombia}

\author[0000-0002-0314-8698]{Antonio Enea Romano}\affiliation{Instituto de Fisica, Universidad de Antioquia, A.A.1226, Medellin, Colombia}

\author[0000-0002-5556-9873]{Rachel Gray}\affiliation{SUPA, University of Glasgow, Glasgow, G12 8QQ, United Kingdom}

\author[0000-0002-6827-9509]{Sergio Vallejo-Pe\~na}\affiliation{Instituto de Fisica, Universidad de Antioquia, A.A.1226, Medellin, Colombia}

\author[0000-0003-2157-4401]{Riccardo Sturani}\affiliation{Universidade Estadual Paulista, 01140-070 São Paulo, Brazil}

\begin{abstract}
Cosmological parameters can be estimated from dark sirens, gravitational waves events without electromagnetic counterpart, by using inference methods which estimate statistically the source redshift from  galaxy catalogs.
We study the effects of applying different apparent magnitude cuts to the GLADE+  catalog, analyzing events from the third LIGO-Virgo-KAGRA gravitational wave transient catalog (GWTC-3), consisting of  42 binary black holes, 3 neutron star–black hole systems and the event GW190425 associated to a binary neutron star coalescence.
We find that the inclusion of fainter galaxies has no significant effects on the joint analysis of almost all the events, while for some single well localized event the effects can be more important, in particular for the event GW190814.
To understand why inclusion of fainter galaxies in general has negligible impact on the estimation of cosmological parameters, we compare the redshift of the events with the redshift distribution of dimmer galaxies, and find that most of them are not located in redshift proximity of the events, explaining the limited impact of their inclusion.
We also find that the larger redshift errors of faint galaxies are not the main factor determining their negligible impact on the estimation of cosmological parameters.
The event GW190814 is an exception, as its distance localization, converted into redshift, overlaps with  fainter galaxies, unlike what happens for other events, which are farther away.
These results show the importance of event localization and catalog depth in determining the impact of catalog completeness on cosmological parameters estimation. To check quantitatively what is the information gained by adding fainter galaxies we have analyzed its impact  on the line of sight prior calculation.
Based on the results for GW190814 we expect that fainter galaxies will play a more significant role in the estimation of cosmological parameters also for the joint analysis of multiple events, when deeper catalogs will be available.
\end{abstract}


\section{Introduction}

The advent of gravitational wave (GW) astronomy has opened a new observational window onto the Universe, providing unique opportunities to probe the physics of compact objects and to address fundamental questions in cosmology \citep{Abbott2016, Abbott2017_GW170817}. One of the most promising applications of GW observations is the use of compact binary coalescences (CBCs) as so-called standard sirens. In these systems, the GW signal encodes a direct measurement of the luminosity distance to the source, independent of the cosmic distance ladder. When an electromagnetic (EM) counterpart is identified, as in the case of GW170817 \citep{Abbott2017_GW170817}, the redshift of the host galaxy can be directly measured, enabling precise estimates of the Hubble constant, $H_{0}$.

However, most GW detections lack an associated EM counterpart. For these dark sirens, the redshift information must be obtained through statistical methods \citep{Schutz1986,Holz:2005df,Fishbach2019}. Specifically, the redshift is inferred from galaxies whose electromagnetic catalog localization overlaps with the GW sky localization. In this approach, the distribution and completeness of galaxies in the catalog play a critical role. Incompleteness or systematics in the galaxy catalog can therefore propagate directly into the inferred cosmological parameters \citep{Gray2020, Gray2023}. In this context, \citet{Datrier2026} investigated a robust statistical approach to characterizing galaxy catalogue completeness through the apparent magnitude completeness limit, and assessed the impact of this treatment on dark siren constraints on the Hubble constant.

Among the available resources, the GLADE+ galaxy catalog \citep{Dalya2021} has been widely adopted in dark siren analyses due to its all sky coverage and multiwavelength photometry. Nevertheless, the catalog suffers from significant incompleteness, particularly in the near-infrared $K$ band\footnote{
K-band is a near-infrared photometric band,
commonly expected to trace stellar mass, as it is less sensitive to star formation which
is traced by the B (blue) band
\cite{Leibler:2010uq,2013ApJ...769...56F}.}.
For example, over 95\% of galaxies in GLADE+ lack a $K$-band magnitude, a limitation that introduces substantial uncertainty in the construction of priors. 
Although the analysis accounts for catalog incompleteness, the absence of $K$-band magnitudes for some galaxies reduces the effective completeness of the catalog and could affect the cosmological inference, particularly if the host galaxy belongs to this population.
This could be especially important for well localized events where catalog properties dominate the prior an effect we investigate and quantify later in this work.

However, a systematic exploration of its impact on line-of-sight (LOS) and on the combined $H_{0}$ constraints for the full set of GWTC-3 \citep{Abbott2023GWTC3} dark sirens remains limited so far. In particular, the dependence of the cosmological results on the level of completeness of the $K$-band magnitudes has not been fully quantified.
Addressing this issue is essential, as future gravitational wave observations with increasing sensitivity will rely heavily on robust statistical dark siren analyses.

A key finding of this study is that the impact of catalog completeness on the LOS prior is governed not only by the number of galaxies included, but also by the precision of their redshift measurements.

In this work, we present a general study of the impact of the \textsc{GLADE+} catalog completeness on dark siren cosmology.
The level of completeness is controlled by  successive cuts in apparent magnitude, corresponding to different relative completeness thresholds (10\%, 30\%, 50\%, 70\%, and 100\%), where the 100\% case refers to using the entire available catalog, i.e. to not applying an apparent magnitude threshold. Note that this definition of completeness does not assume all existing galaxies are detected to some magnitude threshold, but that all the detected galaxies are included in the analysis.
For each threshold, we construct LOS redshift priors and systematically compare them with the \emph{empty-catalogue} case, i.e. to the
case of no galaxy in the catalog.
Our analysis proceeds in four main steps.
First, we assess the incompleteness of the $K$-band magnitudes by determining the fraction of galaxies with and without $K$-band measurements, and identify two distinct subpopulations that differ in their redshift precision.
Second, we analyze the redshift distributions of the GWTC-3 dark siren events and compare them with galaxy-based priors under different completeness assumptions.
Third, we focus on the particularly well localized event GW190814 \citep{Abbott2020GW190814}, which serves as a benchmark case to illustrate the sensitivity of the inferred Hubble constant $H_{0}$ to catalog completeness.
Finally, we extend our analysis to the full GWTC-3 sample of 46 dark sirens, combining the constraints and exploring the correlation between sky-localization area and the magnitude of catalog-induced systematics.

Our results provide a systematic framework to quantify the impact of catalog completeness on LOS and on the robustness of dark siren cosmology. By explicitly comparing different completeness levels, we assess the reliability of current $H_{0}$ measurements and provide guidance for future analyses in the era of advanced GW detectors and deeper galaxy surveys.

\noindent The paper is organized as follows: Section~2 presents the formalism and methodology used to construct the LOS redshift priors. Section~3 presents the main results and their discussion and the conclusions and future prospects are presented in Section~4.

\section{Method}
\label{sec:method}

We infer cosmological parameters from dark sirens by combining the GW luminosity distance posteriors with a LOS built from the \textsc{GLADE+} galaxy catalogue \citep{Dalya2021}. Following \citet[][Sec.~2]{Gray2023}, the LOS prior in a HEALPix (\cite{Gorski:2004by}) pixel $\Omega_i$ is modeled as a mixture of an \emph{in-catalogue} contribution---from galaxies present in the catalogue that pass an apparent magnitude cut in the $K$ band and an \emph{out-of-catalogue} term for galaxies fainter than the adopted apparent magnitude threshold. In the present work completeness is controlled \emph{only} by an apparent magnitude limit $m_{\rm th}(\Omega_i)$;  we do not cut in absolute magnitude, since observational incompleteness is fundamentally determined by the telescope’s flux sensitivity rather than by intrinsic luminosity. Using apparent magnitudes thus provides a realistic way to mimic the observational depth of the survey without introducing cosmology-dependent assumptions through distance estimates.

\subsection{Hierarchical Bayesian framework}

This subsection reviews the formalism adopted for cosmological inference, based on \cite{Gray2020,Gray2023}.

We aim to infer a set of cosmological and population hyper-parameters $\Lambda$ from a catalogue of gravitational wave (GW) data $\{x_{\rm GW}\}$, corresponding to $N_{det}$ detections $\{D_{\rm GW}\}$, which is a binary parameter indicating if an event is detected or not. Within a hierarchical Bayesian framework, the posterior distribution is given by

\begin{widetext}
\begin{equation}
p(\Lambda \mid \{x_{\rm GW}\}, \{D_{\rm GW}\})
\propto
p(\Lambda)\, p(N_{\rm det} \mid \Lambda)
\prod_{i}^{N_{\rm det}}
p(x_{{\rm GW_i}} \mid D_{{\rm GW_i}}, \Lambda).
\label{eq:2.1}
\end{equation}
\end{widetext}

To make the dependence on the source parameters $\theta$
(e.g.\ masses, sky location, inclination, and redshift) explicit,
the likelihood can be written as
\begin{widetext}
\begin{equation}
\begin{aligned}
p(\Lambda \mid \{x_{\rm GW}\}, \{D_{\rm GW}\}, I)
\propto\;&
p(\Lambda \mid I)\,
p(N_{\rm det}\mid \Lambda, I)\,
\left[
\int p(D_{\rm GW}\mid \theta, \Lambda, I)\, p(\theta\mid \Lambda, I)\, d\theta
\right]^{-N_{\rm det}}
\\
&\times
\prod_{i}^{N_{\rm det}}
\int
p(x_{\rm GW_i}\mid \theta,\Lambda,I)\,
p(\theta\mid \Lambda,I)\,
d\theta_i \, .
\end{aligned}
\label{eq:2.2}
\end{equation}
\end{widetext}

The term $I$ encompasses all assumptions that are not explicitly specified in the expression.

GW selection effects are related to the quantity,
\begin{equation}
p(D_{\rm GW}\mid \Lambda,I)
=
\int
p(D_{\rm GW}\mid \theta,\Lambda,I)\,
p(\theta\mid \Lambda,I)\,
d\theta \, ,
\end{equation}
which is the probability of detection for a CBC drawn from a population described by $\Lambda$.


Eq.(\ref{eq:2.2}) can be rewritten to include explicitly the angular dependency, giving
\begin{widetext}
\begin{equation}
\begin{aligned}
p(\Lambda \mid \{x_{\rm GW}\}, \{D_{\rm GW}\}, I)
&\propto p(\Lambda \mid I)\, p(N_{\rm det} \mid \Lambda, I) \\
&\quad \times
\left[
\int \sum_{j}^{N_{\rm pix}}
p(D_{\rm GW} \mid \Omega_j, \theta, \Lambda, I)
\, p(\theta \mid \Omega_j, \Lambda, I)\, d\theta
\right]^{-N_{\rm det}} \\
&\quad \times \prod_{i}^{N_{\rm det}} \left[ \int \sum_{j}^{N_{\rm pix}} p(x_{\rm GW_i}\mid \Omega_j, \theta, \Lambda, I)\, p(\theta\mid \Omega_j, \Lambda, I)\, d\theta \right] .
\end{aligned}
\label{eq:2.3}
\end{equation}
\end{widetext}

\subsection{Introduction of the line-of-sight redshift prior}

Separating also the redshift $z$ from the remaining parameters $\theta = \{z, \theta'\}$, and assuming that the remaining parameters are independent of sky position, the expression becomes

\begin{widetext}
\begin{equation}
\begin{aligned}
p(\Lambda \mid \{x_{\rm GW}\}, \{D_{\rm GW}\}, I)
&\propto p(\Lambda \mid I)\, p(N_{\rm det} \mid \Lambda, I) \\
&\quad \times
\left[
\iint p(D_{\rm GW} \mid z, \theta', \Lambda, I)
\, p(\theta' \mid \Lambda, I)
\sum_{j}^{N_{\rm pix}} p(z \mid \Omega_j, \Lambda, I)\, d\theta' dz
\right]^{-N_{\rm det}} \\
&\quad \times
\prod_{i}^{N_{\rm det}}
\iint \sum_{j}^{N_{\rm pix}}
p(x_{{\rm GW_i}} \mid \Omega_j, z, \theta', \Lambda, I)
\, p(\theta' \mid \Lambda, I)
\, p(z \mid \Omega_j, \Lambda, I)\, d\theta' dz \,,
\end{aligned}
\label{eq:2.4}
\end{equation}
\end{widetext}
where sky angle marginalization has been replaced by a sum over pixels.

The LOS redshift prior $p(z \mid \Omega_j, \Lambda, I)$ encodes information about the galaxy distribution in each pixel $\Omega_j$.
The LOS can be decomposed in into an in-catalogue and out-of-catalogue contributions,
the latter taking into account galaxies missing from the catalog:

\begin{widetext}
\begin{equation}
\begin{aligned}
p(z \mid \Omega_i, \Lambda, s, I)
&= \iint \sum_{g \in \{G, \bar{G}\}}
p(z, M, m, g \mid \Omega_i, \Lambda, s, I)\, dM\, dm \\
&= p(G \mid \Omega_i, \Lambda, s, I)
\iint p(z, M, m \mid G, \Omega_i, \Lambda, s, I)\, dM\, dm \\
&\quad + p(\bar{G} \mid \Omega_i, \Lambda, s, I)
\iint p(z, M, m \mid \bar{G}, \Omega_i, \Lambda, s, I)\, dM\, dm \, ,
\end{aligned}
\label{eq:los_full_decomposition}
\end{equation}
\end{widetext}
where $z$ is the redshift of the source, $M$ and $m$ are the absolute and apparent magnitudes of the host galaxy, and $g$ is a binary variable indicating whether the host galaxy is contained within the catalogue ($G$) or not  ($\bar{G}$). The term $\Omega_i$ denotes the sky direction (pixel), $\Lambda$ represents the set of cosmological and population hyper-parameters, $s$ indicates the presence of a GW source, and $I$ encodes any additional assumptions.
Following \cite{Gray2023}, the term $p(z, M, m \mid G, \Omega_i, \Lambda, s, I)$
is the in-catalog contribution, corresponding to galaxies which are selected by the condition $m\leq m_{th}(\Omega_i)$ , where $m_{th}$ is the magnitude threshold for.

Conversely, $p(z, M, m \mid \bar{G}, \Omega_i, \Lambda, s, I)$
describes the out catalog contribution,  corresponding to the galaxies satisfying the condition $m > m_{th}(\Omega_i)$.

Combining the in-catalogue and out-of-catalogue contributions, the final LOS redshift prior takes the form \cite{Gray2023}

\begin{widetext}
\begin{equation}
\begin{aligned}
p(z \mid \Omega_i, \Lambda, s, I)
&=
\frac{1}{p(s \mid \Omega_i, \Lambda, I)}
\Bigg[
p(G \mid \Omega_i, \Lambda, I)
\frac{1}{N_{\rm gal}(\Omega_i)}
\sum_{k=1}^{N_{\rm gal}(\Omega_i)}
p(z \mid \hat{z}_k)\,
p(s \mid z, M(z, \hat{m}_k, \Lambda), \Lambda, I)
\\
&\quad +
\Theta[z_{\rm cut}-z]
\int_{M(z,m_{\rm th}(\Omega_i),\Lambda)}^{M_{\rm max}(H_0)}
p(z, M \mid \Lambda, I)\, p(s \mid z, M, \Lambda, I)\, dM
\\
&\quad +
\Theta[z-z_{\rm cut}]
\int_{M_{\rm min}(H_0)}^{M_{\rm max}(H_0)}
p(z, M \mid \Lambda, I)\, p(s \mid z, M, \Lambda, I)\, dM
\Bigg] \, .
\end{aligned}
\label{eq:los_final}
\end{equation}
\end{widetext}
The first line shows the in-catalogue contribution to the LOS prior which is given by a discrete sum over galaxies, where $\hat{z}_k$ and $\hat{m}_k$ denote the observed redshift and apparent magnitude of the $k$-th galaxy in pixel $\Omega_i$.

The second and third lines correspond to the out-catalogue contribution, where $z_{\rm{cut}}$ is a  redshift threshold beyond which the catalog data is not considered reliable, and is hence not included in the LOS calculation.

\subsection[Controlling completeness with the apparent magnitude threshold in K]%
{Controlling completeness with the apparent magnitude threshold}
\label{subsec:completeness_mth}

We construct per-pixel maps of the \emph{apparent}-magnitude threshold $m_{\rm th}(\Omega_i)$ in the $K$ band at completeness percentiles $p \in \{10,30,50,100\}$. 
Operationally, for each pixel $\Omega_i$, we consider the empirical distribution of $K$-band apparent magnitudes of the catalogue galaxies and define $m_{\rm th}^{(p)}(\Omega_i)$ as the $p$th percentile of that distribution (e.g., $p=50$ corresponds to the median).

Because apparent magnitudes are inversely ordered with brightness, \emph{lower} percentiles correspond to \emph{brighter} thresholds (i.e., stricter selection), while \emph{higher} percentiles correspond to \emph{fainter} thresholds (i.e., more permissive selection). 
This construction ensures that catalogue completeness is controlled \emph{exclusively} through an apparent magnitude cut that adapts to the local depth of the survey across the sky.

All LOS priors used in this work are constructed using this apparent magnitude based definition of completeness. By default, the \texttt{GWCOSMO} pipeline \cite{Gray2020,Gray:2021sew} used in GWTC-3 adopts $m_{\rm th}(\Omega_i)$ at the 50th percentile in each pixel. Here, we explicitly vary this threshold across the range $10$--$100$\% to quantify how different levels of catalogue completeness impact the LOS prior and, consequently, the inferred cosmological parameters.

Varying $m_{\rm th}(\Omega_i)$ directly modifies the balance between the in-catalogue and out-of-catalogue contributions to the LOS prior (Eq.~\ref{eq:los_final}). Lower percentiles include only the brightest galaxies, leading to stronger incompleteness and a larger contribution from the out-of-catalogue term. Conversely, higher percentiles progressively incorporate fainter galaxies, increasing $N_{\rm gal}(\Omega_i)$ and reducing the relative importance of the out-of-catalogue component.
In this sense, increasing $m_{\rm th}(\Omega_i)$ from the 10th to the 100th percentile changes the out-of-catalogue and the in-catalogue contributions, thereby reshaping the LOS redshift prior $p(z \mid \Omega_i)$.


\subsection{Pipeline with \texttt{gwcosmo} and event sample}
\label{subsec:pipeline}

We use \texttt{GWCOSMO} to construct LOS priors $p_t(z)$ from the GLADE+ catalogue in the $K$ band with resolution\footnote{The HEALPix parameter $n_{\rm side}$ sets the
angular resolution of the LOS prior on the sky.} $n_\mathrm{side}=64$, for a set of percentile thresholds $t \in \{10,30,50,100\}$. Here, $p_t(z)$ denotes the LOS prior corresponding to the apparent magnitude threshold defined at percentile $t$ in each pixel. We also define an empty-catalogue prior $p_0(z)$, corresponding to the limiting case in which no galaxy information is used.

For each event $i$, the cosmological posterior for $H_0$ is obtained by evaluating Eq.~(4), where 
the GW likelihood and the LOS redshift prior $p_t(z)$ are evaluated together, with the likelihood marginalised over sky location, redshift, and other source parameters.

Our sample comprises 46 GWTC-3 dark siren events (excluding GW170817, which is treated as a bright siren). For each event, we evaluate Eq.~(4) using LOS priors $p_t(z)$ at thresholds $t \in \{10,30,50,100\}$, as well as the empty-catalogue prior $p_0(z)$. We then compare the resulting single-event constraints using the best-fit values (i.e., the parameter values that maximise each posterior), their $68\%$ and $95\%$ credible intervals, and the corresponding relative uncertainties. Combined constraints are obtained by multiplying the independent single-event posteriors.


At fixed $n_{\mathrm{side}}$ and population model, varying only the apparent magnitude threshold $m_{\rm th}^{(t)}(\Omega_i)$ allows us to assess the relative effect of galaxy catalogue completeness on the LOS prior $p_t(z)$ and, through Eq. (4), on the inferred value of $H_0$, while keeping in mind that the fixed population model itself provides a lot cosmological information.

\section{Results and Discussion}

\subsection{Multi-event $H_0$ estimation for 42 BBHs}

\begin{figure*}[htbp]
  \centering
  \includegraphics[width=0.90\textwidth]{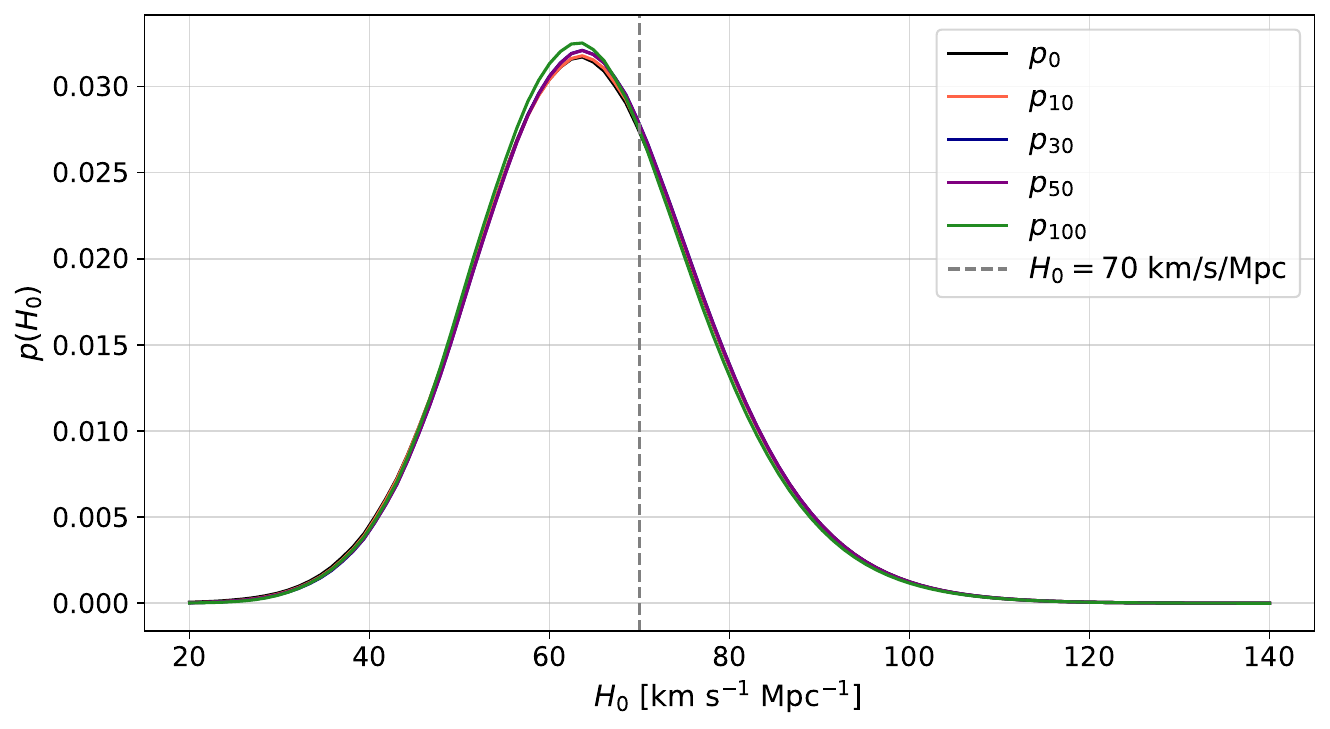}
  \caption{Joint $H_0$ posterior from 42 BBHs for completeness percentiles 10\% (red), 30\% (blue), 50\% (darkblue), and 100\% (green). The vertical dashed line marks $H_0 = 70~\mathrm{km~s^{-1}~Mpc^{-1}}$.}
  \label{fig:H0_multievent}
\end{figure*}

The results of the joint analysis of GWTC-3 42 BBHs are shown in Fig.~\ref{fig:H0_multievent} and Table~\ref{tab:H0_multievent_table}. \footnote{We use the definition $\mathrm{Rel.\ Err.}= \frac{\text{Upper}-\text{Lower}}{\text{Mean}} \times 100$ uniformly in all tables.}

The LOS priors were computed from the GLADE+ galaxy catalogue with
$n_{\rm side}=64$, and completeness percentiles $0\%, 10\%, 30\%$,
$50\%$, and $100\%$. The BBH population model adopted in this analysis was the
BBH--power-law--Gaussian model, following the population model used in
the GWTC-3 cosmological analysis \cite{GWTC3}. For the
redshift evolution of the merger rate, we assumed a \cite{Madau:2014bja}
parametrization, with $\gamma=4.59$, $z_p=2.47$,
and $k=2.86$

\begin{table*}[t]
\centering
\caption{Best--fit values and 68\% and 95\% confidence intervals for the joint $H_0$ posterior distribution using different completeness percentiles of the catalog (42 BBHs). We report the best--fit estimate, the 68\% and 95\% credible intervals, and the relative errors (width of the interval divided by the best--fit value).}
\label{tab:H0_multievent_table}
\begin{tabular}{lccccc}
\hline
Percentile & $\boldsymbol{H_0}$ [km s$^{-1}$ Mpc$^{-1}$] & 68\% CI & 95\% CI & Rel.\ Err.\ (68\%) & Rel.\ Err.\ (95\%) \\
\hline
0   & $63.64^{+13.20}_{-12.24}$ 
& (51.40, 76.84) & (39.80, 91.83)
& 39.99\% & 81.76\% \\

10  & $63.64^{+13.18}_{-12.17}$ 
& (51.47, 76.82) & (40.11, 91.71])
& 39.84\% & 81.09\% \\

30  & $63.64^{+13.23}_{-11.94}$ 
& (51.70, 76.87) & (40.39, 91.73) 
& 39.55\% & 80.68\% \\

50  & $63.64^{+13.24}_{-11.96}$ 
& (51.68, 76.88) & (40.31, 91.79) 
& 39.60\% & 80.91\% \\

100 & $63.64^{+12.76}_{-12.15}$ 
& (51.49, 76.40) & (40.33, 91.36) 
& 39.14\% & 80.18\% \\
\hline

\end{tabular}
\end{table*}

\subsection{Combining 42 BBHs + 3 NSBHs + 1 BNS single events posteriors}

\begin{figure*}[htbp]
  \centering
  \includegraphics[width=0.90\textwidth]{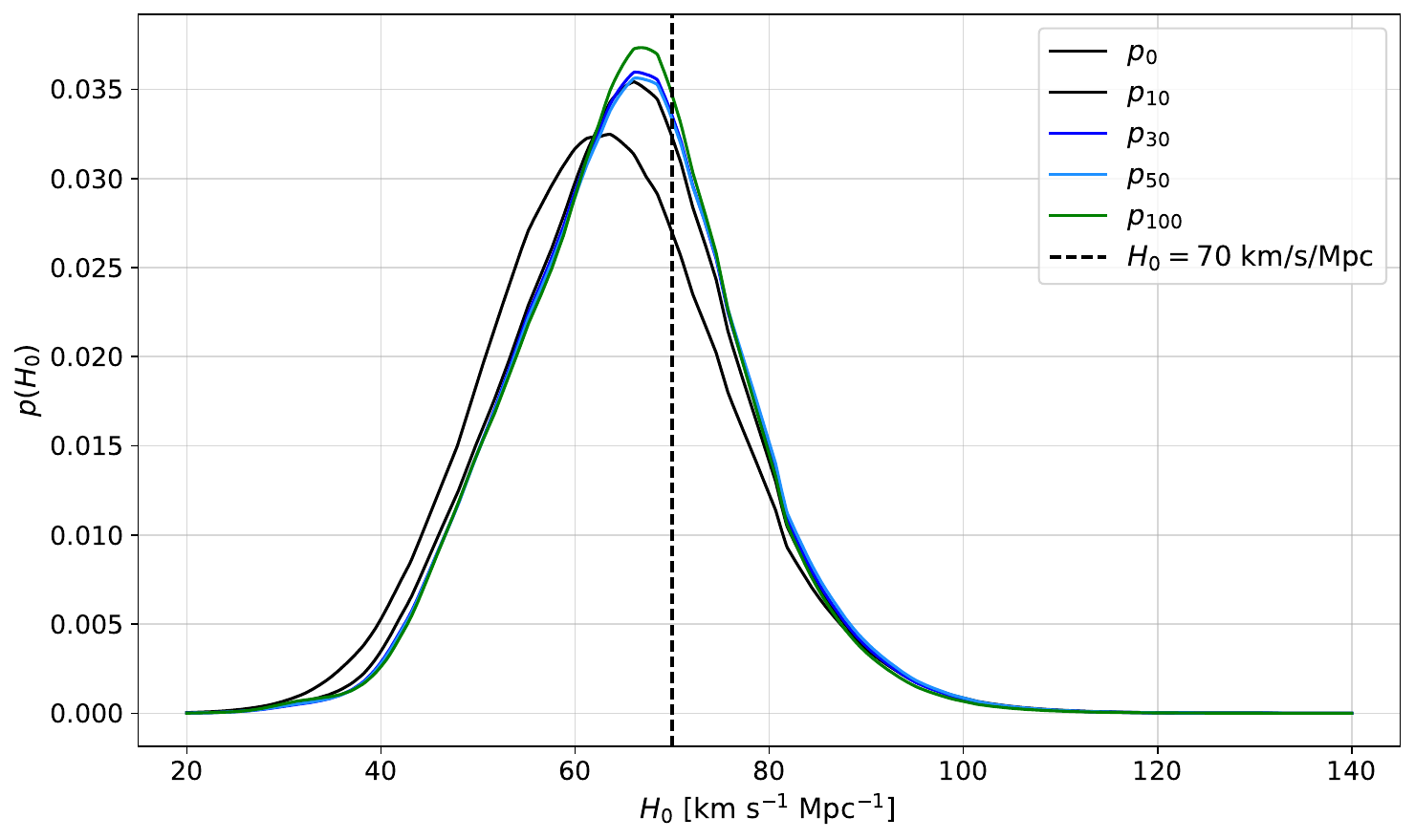}
  \caption{Combined $H_0$ posterior for 46 events (42 BBHs + 3 NSBHs + GW190425) for different value of the completeness percentile. Posteriors are obtained by multiplying the single-event posteriors computed with LOS redshift priors at each percentile. The vertical dashed line marks $H_0 = 70~\mathrm{km~s^{-1}~Mpc^{-1}}$.}
  \label{fig:H0_combined}
\end{figure*}

In Fig.~\ref{fig:H0_combined} and Table~\ref{tab:H0_combined_table} we show the results of the combination of single event posteriors, which are in agreement with  the results of the joint analysis, which 
analyzes all events simultaneously.

\begin{table*}[htbp]
\centering
\caption{Best--fit estimates of the combined $H_0$ posterior using different catalog--completeness percentiles (46 events). We report the best--fit value with 68\% errors, the central 68\% and 95\% credible intervals, and the relative errors (interval width divided by the best--fit value).}
\label{tab:H0_combined_table}
\begin{tabular}{lccccc}
\hline
Percentile & $\boldsymbol{H_0}$ [km s$^{-1}$ Mpc$^{-1}$] & 68\% CI & 95\% CI & Rel.\ Err.\ (68\%) & Rel.\ Err.\ (95\%) \\
\hline

0   & $63.60^{+12.08}_{-12.56}$
& (51.04, 75.68) & (39.77, 89.93)
& 38.74\%  & 78.86\% \\

10 & $66.01^{+10.55}_{-12.88}$ & (53.13, 76.56) & (42.11, 89.84) & 35.50\% & 72.30\% \\ 30 & $66.25^{+10.57}_{-12.48}$ & (53.77, 76.82) & (43.03, 89.81) & 34.79\% & 70.61\% \\ 50 & $66.37^{+10.74}_{-12.54}$ & (53.82, 77.10) & (43.08, 90.23) & 35.08\% & 71.04\% \\ 100 & $66.85^{+9.61}_{-13.00}$ & (53.85, 76.46) & (43.07, 88.90) & 33.82\% & 68.57\% \\ \hline \end{tabular} \end{table*}

\subsubsection{Detailed analysis of GW190814: impact of catalogue completeness}

The event GW190814 is located at a fiducial distance of $\sim 240$~Mpc and has a sky localization area of $18.5~\mathrm{deg}^{2}$ at the 90\% HDI \citealt{Abbott2020GW190814}). 

\begin{figure}[t]
    \centering
    \includegraphics[width=0.48\textwidth]{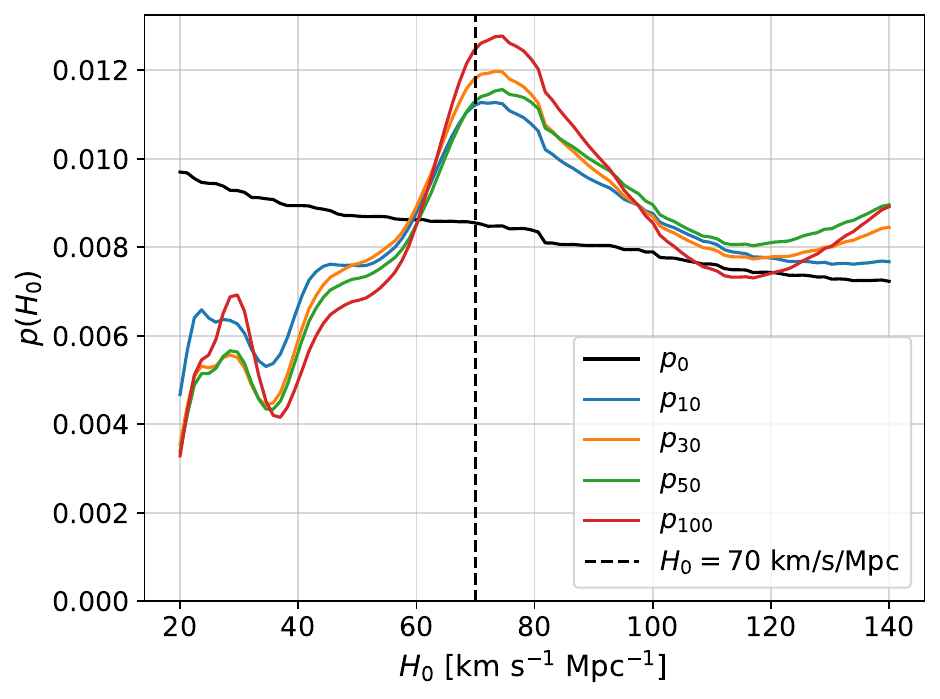}
    \caption{Posterior distributions of $H_{0}$ for GW190814 using LOS priors computed at 
    different completeness percentiles, and computed with the empty catalogue LOS. The vertical dashed line indicates $H_{0}=70~\mathrm{km\,s^{-1}\,Mpc^{-1}}$.}
    \label{fig:gw190814_post}
\end{figure}

Fig.~\ref{fig:gw190814_post} shows the $H_{0}$ posteriors obtained for GW190814 when using the LOS constructed with different apparent magnitude cuts corresponding to the percentiles $\{0,10,30,50,100\}$. Table~\ref{tab:gw190814_map_ci} reports the corresponding best  values and confidence intervals.

The event GW190814 differs from  the majority of other events, because its fiducial redshift is within the range well covered by $K$-band part of the catalogue, as shown in Fig.~\ref{fig:redshift_subgroups_gw}, implying a more significant effect on the   LOS and the $H_0$ posterior.
Including fainter galaxies, higher redshift objects are included, shifting the posterior distribution towards larger values of $H_{0}$. This is visible in both Fig.~\ref{fig:gw190814_post} and Table~\ref{tab:gw190814_map_ci}.

While noticeable, this shift is not very pronounced, because  fainter galaxies tend to have larger redshift uncertainties $\sigma_{z}$ which reduce the informativeness of the in-catalog component of the LOS prior. 
We used the sky localization map released by the gravitational wave transient LIGO--Virgo--KAGRA Collaboration catalog (IGWN--GWTC2.1 \cite{Abbott2024GWTC21}). 

\begin{figure}[htbp]
    \centering
    \includegraphics[width=0.48\textwidth]{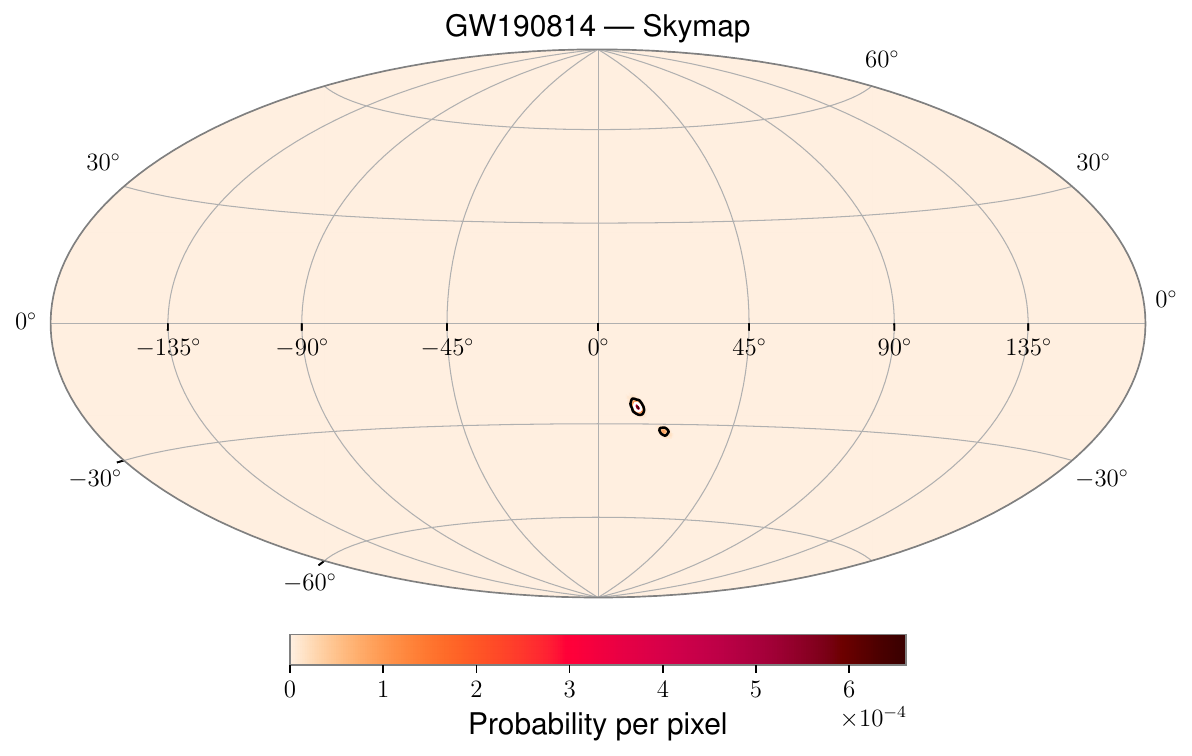}
    \caption{Sky localization of GW190814 from the IGWN--GWTC2.1 data release.
The sky map was generated from the posterior samples using
\texttt{ligo-skymap-from-samples} from the \texttt{ligo.skymap} package
\citep{Singer2016GoingDistance,Singer2016Supplement}.
The probability map is shown in equatorial coordinates with a Mollweide
projection. The colour scale encodes the posterior probability density per
pixel, and the 90\% credible region is highlighted with a contour.}
    \label{fig:gw190814_skymap}
\end{figure}

The $90\%$ credible region, shown in Fig.~\ref{fig:gw190814_skymap}, covers $18.5~\mathrm{deg}^{2}$. The good sky localization of GW190814 plays an important role in increasing the difference with respect to the empty catalogue LOS prior, when including fainter galaxies in the LOS calculation. 

\begin{table*}[htbp]
    \centering
    \caption{Best--fit values of $H_{0}$ for GW190814 at different completeness percentiles, 
    with $68\%$ and $95\%$ central credible intervals and their relative errors 
    (computed as $(\mathrm{upper}-\mathrm{lower})/\mathrm{mean}\times 100$).}
    \label{tab:gw190814_map_ci}
    \begin{tabular}{lccccc}
        \hline
        Percentile & $\boldsymbol{H_0}$ [km s$^{-1}$ Mpc$^{-1}$] & 68\% CI &  95\% CI & Rel.\ Err.\ (68\%)  & Rel.\ Err.\ (95\%) \\
        \hline
        0   & $20.00^{+98.37}_{-16.98}$ 
        & (38.98, 118.37) & (22.57, 136.99)
        & 406.94\%  & 570.11\% \\

        10  & $73.33^{+45.80}_{-28.77}$ 
        & (44.56, 119.13) & (23.14, 136.74) 
        & 101.69\%  & 154.91\% \\

        30  & $73.33^{+46.75}_{-25.73}$ 
        & (47.60, 120.18) & (24.16, 137.03) 
        & 98.97\% & 153.92\% \\

        50  & $74.55^{+46.66}_{-26.38}$ 
        & (48.17, 121.21) & (24.34, 137.20) 
        & 97.98\% & 151.39\% \\

        100 & $74.55^{+45.77}_{-26.73}$ 
        & (47.82, 120.32) & (24.18, 137.18) 
        & 97.26\% & 151.58\% \\
        \hline
    \end{tabular}
\end{table*}

\subsection[Single-event H0 posteriors across completeness percentiles]%
{Single-event $H_0$ posteriors across completeness percentiles}

\label{sec:results_individual_events}

\begin{figure*}[t]
  \centering
  \includegraphics[width=\textwidth]{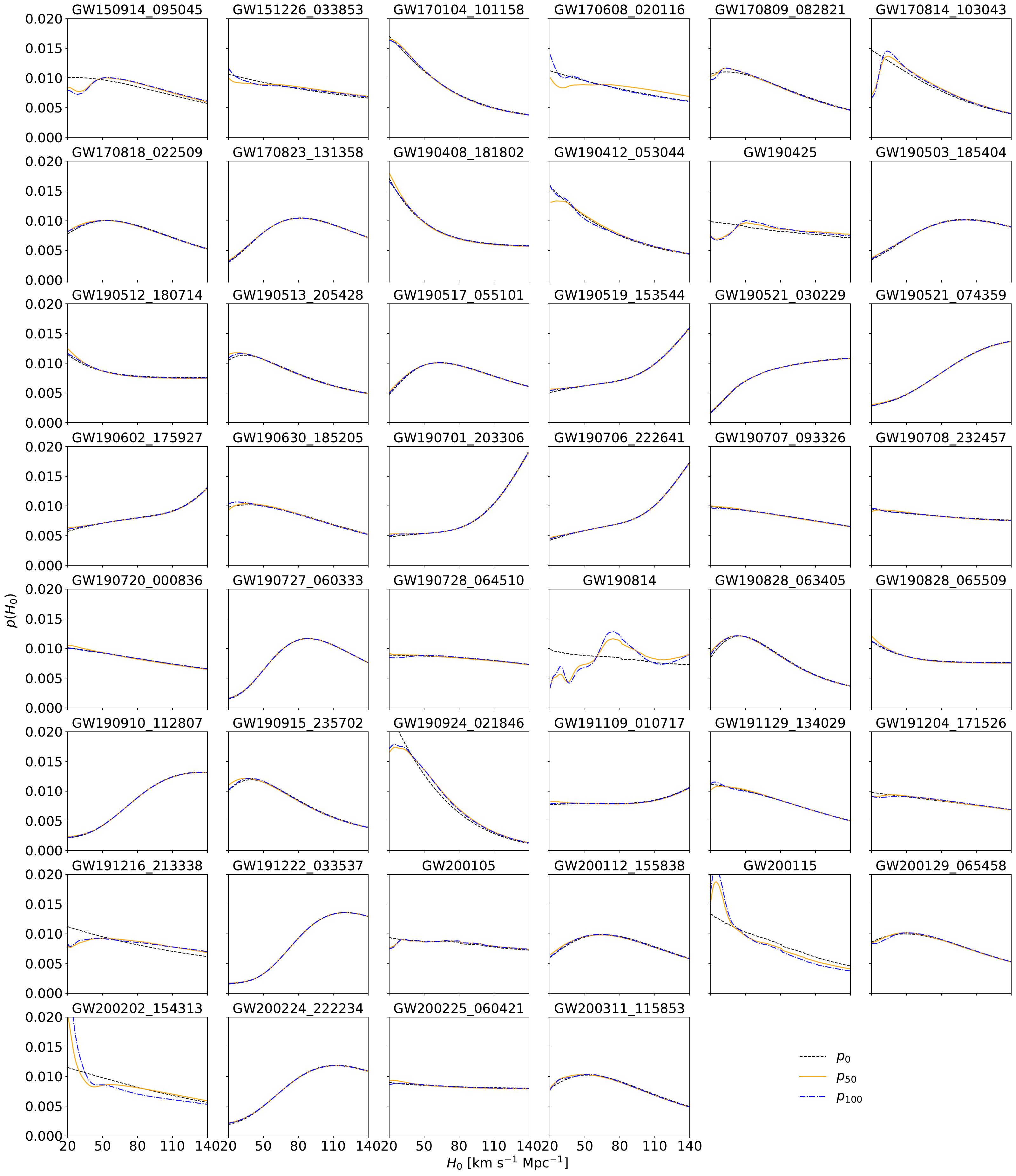}
  \caption{$H_0$ posteriors for 42 BBHs, 3 NSBHs and GW190425 for different LOS obtained for different values of the completeness percentile. Each panel shows one event; curves of different colors correspond to different percentiles.}
\label{fig:H0_individual}
\end{figure*}

Fig.~\ref{fig:H0_individual} compares the $H_0$ posterior for each GWTC-3 event obtained imposing different apparent magnitude cuts. We analyze 42 BBHs,  3 NSBHs and GW190425 (BNS), computing the LOS prior with a healpix resolution $n_\mathrm{side}=64$, using the GLADE+ galaxy catalogue ($K$ band). For BBHs we adopt the \texttt{BBH‑powerlaw‑gaussian} mass model and  for NSBHs (e.g., GW190814) we use the \texttt{NSBH‑powerlaw‑gaussian} model,
according to the findings of \cite{KAGRA:2021duu}, 
with neutron‑star masses in the range  $M_\mathrm{NS}\in[1.0,3.0]\,M_\odot$ and zero NS spin ($a_\mathrm{NS}=0$). In all runs we fix the population–model parameters (mass distribution, redshift evolution) while fitting $H_0$. We analyze events with SNR $\ge 11$. The apparent magnitude $m_{th}$  thresholds correspond to the percentiles 0, 10, 30, 50, 100.

We find that for most of the events the different apparent magnitude thresholds have negligible effects on the  $H_{0}$ posteriors.
To understand why fainter galaxies have a negligible impact on the estimation of cosmological parameters, we compare the redshift of the events with the redshift distribution of fainter galaxies, and find that most of them are not located in redshift proximity of the events, explaining the limited impact of including them, see e.g. Fig.~\ref{fig:redshift_subgroups_gw}.
We also compare the redshift uncertainty of bright and faint galaxies, finding that the latter have on average larger redshift uncertainties, further explaining why they have negligible impact on the cosmological parameters estimation.

For the event GW190814 the fainter galaxies are located in redshift proximity of the event, and consequently have a non negligible effects on the LOS, and the cosmological parameters estimation.

\subsection{K-band completeness and redshift error distribution}
\label{sec:results_kband}

The LOS redshift priors used in this work were constructed using the $K$-band magnitudes from the GLADE+ galaxy catalogue. However, as shown in Table~\ref{tab:missing_mags}, this band is severely incomplete: 95.4\% of galaxies in the catalog lack a $K$-band magnitude. This level of incompleteness is far greater than in other available bands such as $B$, $W1$, or $bJ$ \footnote{The W1 band is the first infrared band of the WISE survey, centered at approximately $3.4\,\mu{\rm m}$ \citep{Wright:2010qw}, while the $b_J$ band is a blue optical photographic band, broadly similar to the standard $B$ band and widely used in galaxy redshift surveys \citep{Norberg:2001br,Colless:2001gk}.}. 


\begin{table}[htbp]
\centering
\caption{Fraction of galaxies in GLADE+ without a measured magnitude in each photometric band.}
\label{tab:missing_mags}
\begin{tabular}{lcc}
\hline
Band & Galaxies without mag. & Percentage (\%) \\
\hline
K   & 20,869,685 & 95.4163 \\
B   &   513,743 &  2.3488 \\
W1  &   101,461 &  0.4639 \\
bJ  & 1,137,107 &  5.1989 \\
\hline
\end{tabular}
\end{table}

\begin{figure*}[htbp]
\centering
\includegraphics[width=0.85\textwidth]{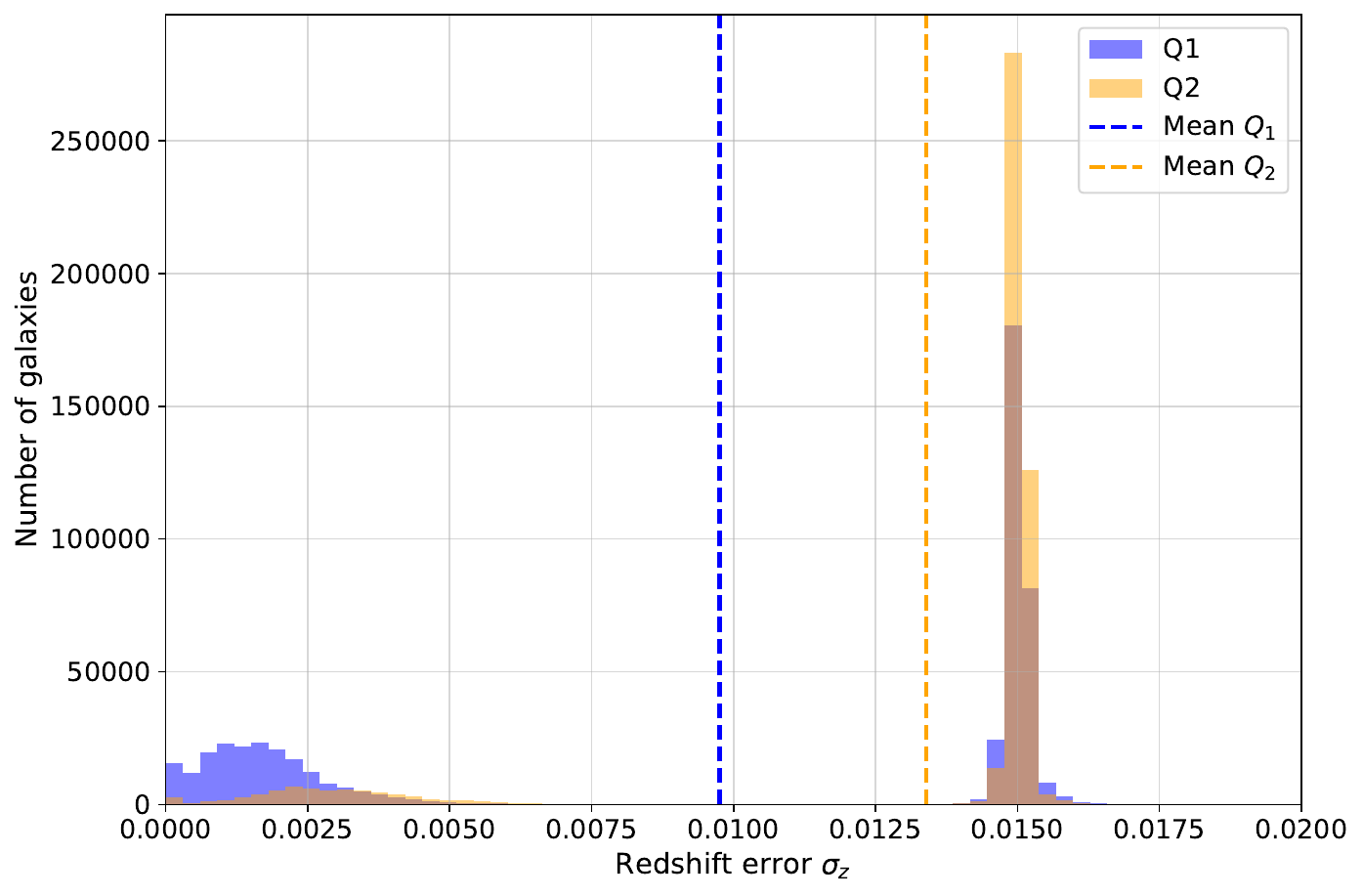}
\caption{Distribution of the redshift error ($\sigma_z$) for galaxies in the brighter half (Q1) and fainter half (Q2) groups of the GLADE+ $K$-band. The  plot shows the histograms of $\sigma_z$ with vertical dashed lines corresponding to the mean value of each group. Brighter galaxies  have a smaller average redshift uncertainty. }
\label{fig:sigmaz_Q1Q2}
\end{figure*}

To assess the quality of the available $K$-band subset, we examined the distribution of redshift errors ($\sigma_z$) for galaxies with measured $K$-band magnitudes. As a first step, the catalog was divided into two groups: galaxies brighter than the 50th percentile magnitude ($m_K \leq 13.523$) and galaxies fainter than this limit ($m_K > 13.523$). In the following, these two groups are denoted as Q1 (brighter half) and Q2 (fainter half), respectively. 
The distributions of $\sigma_z$ for the two groups can be seen in Fig.~\ref{fig:sigmaz_Q1Q2}. Fainter galaxies tend to have larger $\sigma_z$ values, with the mean redshift error increasing from 0.009757 in the brighter half to 0.013393 in the fainter half. This trend is consistent with the expectation that fainter galaxies, often located at larger distances, have less precise redshift measurements. Summary statistics for the two groups are listed in Table~\ref{tab:sigmaz_Q1Q2}. 


\begin{table*}[htbp]
\centering
\caption{Statistical summary of the redshift error ($\sigma_z$) for the two groups of  GLADE+ K=band galaxies, defined by apparent magnitude.}
\label{tab:sigmaz_Q1Q2}
\begin{tabular}{l r r r c r c r}
\hline
Group & $N$ & Mean $\sigma_z$ & Median $\sigma_z$ & 68\% CI & 95\% CI \\
\hline
Q1 (Brighter half) & 501{,}700 & 0.009757 & 0.014830 & [0.001333, 0.015101] & [0.000161, 0.015389] \\
Q2 (Fainter half)  & 500{,}855 & 0.013393 & 0.014977 & [0.014748, 0.015118]  & [0.001804, 0.015272]  \\
\hline
\end{tabular}
\end{table*}

\begin{figure*}[htbp]
\centering
\includegraphics[width=0.9\textwidth]{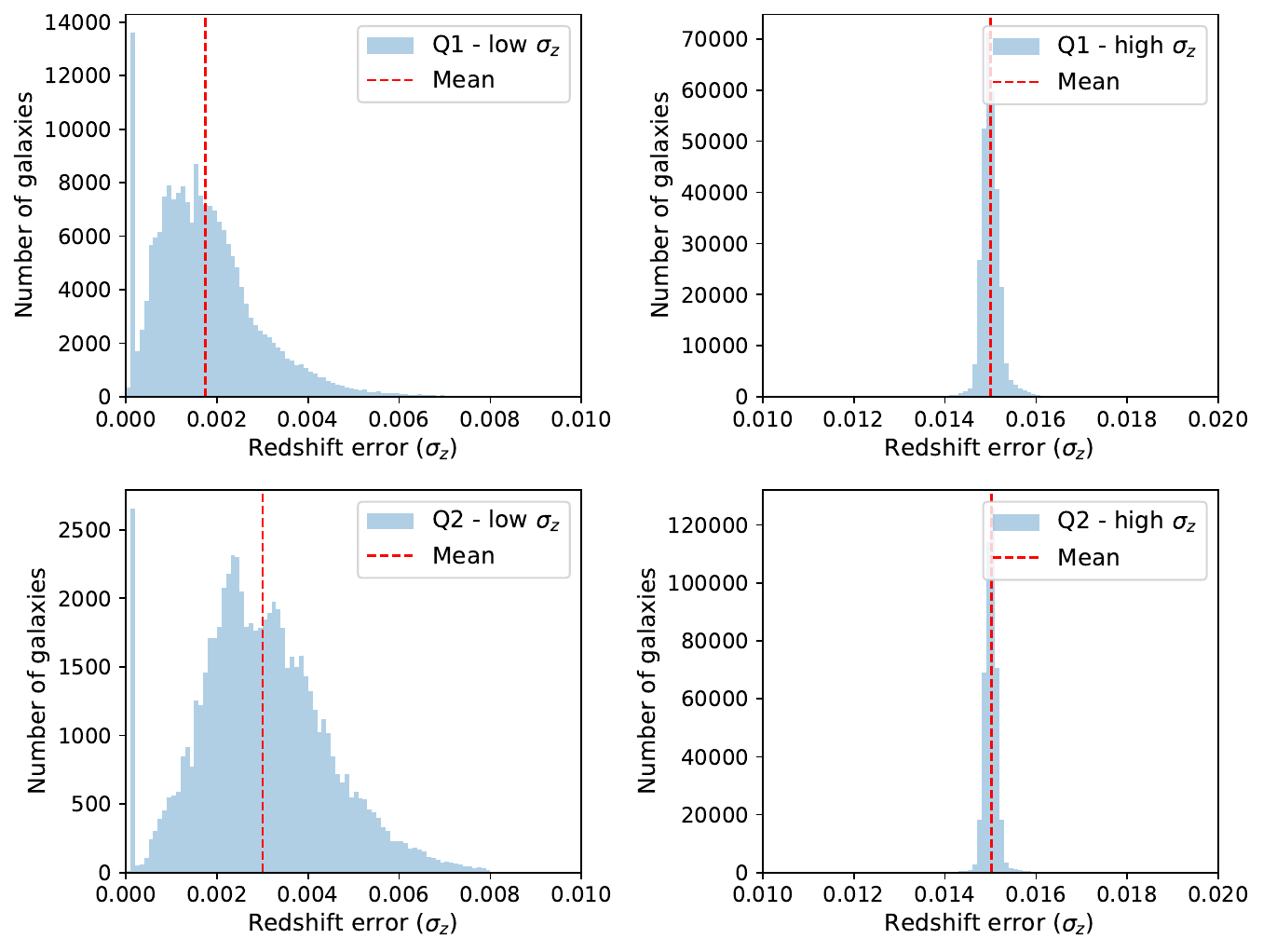}
\caption{Distributions of redshift errors ($\sigma_z$)  of GLADE+ galaxies with $K$-band magnitudes in each of the four subgroups Q1-low $\sigma_z$, Q1-high $\sigma_z$, Q2-low $\sigma_z$, and Q2-high $\sigma_z$. 
To obtain these four subgroups, the sample is divided into brighter  ($m_K \leq 13.523$) and fainter ($m_K > 13.523$) galaxies, and each of these groups is further divided into two subgroups according to the bimodality of their $\sigma_z$ distributions. 
The low-$\sigma_z$ subgroups correspond to a minority of galaxies with very small redshift errors but large statistical scatter, while the high-$\sigma_z$  subgroups dominate in number and are tightly clustered around $\sigma_z \sim 0.015$. 
Red dashed lines indicate the mean values.}
\label{fig:quartiles}
\end{figure*}

For a more detailed analysis, each of these two groups was further divided into two subgroups according to the bimodal distribution of $\sigma_z$ in Fig.~\ref{fig:sigmaz_Q1Q2}. As can be seen in this figure, each histogram shows two distinct regimes: a small fraction of galaxies with very low $\sigma_z$ values (typically $\lesssim 0.005$) and a dominant population clustered around $\sigma_z \sim 0.015$. This yields four subgroups in total: Q1-low $\sigma_z$, Q1-high $\sigma_z$, Q2-low $\sigma_z$, and Q2-high $\sigma_z$, and their corresponding redshift error distributions are shown in Fig.~\ref{fig:quartiles}.

The statistical properties of these four subgroups are summarized in Table~\ref{tab:quartiles}. 
Importantly, the low-$\sigma_z$ groups in Q1 and Q2 show small mean errors but very large relative uncertainties, indicating statistical instability. In contrast, the high-$\sigma_z$ groups contain the majority of galaxies and exhibit extremely stable distributions centered at $\sigma_z \sim 0.015$, with relative errors of only a few percent.

\begin{table*}[t]
\centering
\caption{Statistical summary of the redshift error ($\sigma_z$) distributions for the four subgroups Q1-low $\sigma_z$, Q1-high $\sigma_z$, Q2-low $\sigma_z$, and Q2-high $\sigma_z$, which were defined by  GLADE+ K-band galaxies brightness and bimodality of their $\sigma_z$ distributions.}
\label{tab:quartiles}
\begin{tabular}{lccccccc}
\hline
Group & $N$ & Mean $\sigma_z$ & Median $\sigma_z$ & 68\% CI & 95\% CI \\
\hline
Q1 - low $\sigma_z$  & 198,608 & 0.00175 & 0.00159 & [0.00067, 0.00275] & [0.00015, 0.00450]  \\
Q1 - high $\sigma_z$ & 303,092 & 0.01500 & 0.01498 & [0.01482, 0.01518] & [0.01465, 0.01554] \\
Q2 - low $\sigma_z$  & 67,968 & 0.00300 & 0.00290 & [0.00169, 0.00437] & [0.00015, 0.00614] \\
Q2 - high $\sigma_z$ & 432,887 & 0.01502 & 0.01501 & [0.01487, 0.01513] & [0.01474, 0.01529] \\
\hline
\end{tabular}
\end{table*}


Crucially, as seen in Tables~\ref{tab:sigmaz_Q1Q2}--\ref{tab:quartiles}, the galaxies that enter when the completeness percentile is increased (i.e., the faint tail) have redshift errors that are slightly larger yet very similar to one another, clustering around $\sigma_z \simeq 0.015$ . In the LOS construction, each galaxy contributes with a truncated-Gaussian distribution $p(z \mid \hat z_k)$ (see Eq. {\ref{eq:los_final}}). 
Therefore, adding  fainter galaxies with   high $\sigma_z$ is not adding significant information to the LOS.
In practice, this means that the additional galaxies contribute little new information: their poorly constrained redshifts broaden $p(z \mid \hat z_k)$ in a way that resembles the out-of-catalogue component. 


To quantify the contribution of the different redshift measurement types to the bimodal structure observed in the $\sigma_z$ distribution (see Fig.~\ref{fig:sigmaz_Q1Q2}), we examined the type of redshift measurement associated with galaxies in each of the four subsamples: Q1--low $\sigma_z$, Q1--high $\sigma_z$, Q2--low $\sigma_z$, and Q2--high $\sigma_z$. The results are summarized in Tables~\ref{tab:redshift_type_q1_q2} and~\ref{tab:redshift_types_subgroups}. The two clusters naturally reflect the different redshift measurement populations: galaxies in the high-$\sigma_z$ subsamples of both Q1 and Q2 are overwhelmingly dominated by photometric redshift estimates, whereas galaxies in the low-$\sigma_z$ subsamples are primarily associated with spectroscopic redshifts.

\begin{table*}
\centering
\begin{tabular}{lcc}
\hline
\textbf{Redshift type} & \textbf{Q1 (\%)} & \textbf{Q2 (\%)} \\
\hline
Photometric    & 61.748 & 87.388 \\
Spectroscopic  & 38.247 & 12.611 \\
\hline
\end{tabular}
\caption{Fraction of photometric and spectroscopic redshifts of groups Q1 and Q2.}
\label{tab:redshift_type_q1_q2}
\end{table*}

\begin{table*}
\centering
\caption{Fraction of photometric and spectroscopic redshifts for each subgroups.}
\begin{tabular}{lcccc}
\hline
\textbf{Redshift type} & \textbf{Q1 low $\sigma_z$ (\%)} & \textbf{Q1 high $\sigma_z$ (\%)} & \textbf{Q2 low $\sigma_z$ (\%)} & \textbf{Q2 high $\sigma_z$ (\%)} \\
\hline
Photometric   & 2.956 & 99.993 & 3.328 & 99.961 \\
Spectroscopic & 97.032 & 0.0068 & 96.669 & 0.039 \\
\hline
\end{tabular}
\label{tab:redshift_types_subgroups}
\end{table*}

\subsubsection[Redshift distributions of the sigma-z subgroups]
{Redshift distributions of the four GLADE+ catalogue K-band subgroups}

In addition to the redshift error distributions, we also investigated the redshift distributions of the galaxies in the four GLADE+ catalogue K-band subgroups Q1-low $\sigma_z$, Q1-high $\sigma_z$, Q2-low $\sigma_z$, and Q2-high $\sigma_z$  (see Fig.~\ref{fig:quartiles}). This complementary view offers insight into the statistical properties of the galaxies' redshift for the four subgroups.

\begin{figure*}[htbp]
    \centering
    \includegraphics[width=0.95\textwidth]{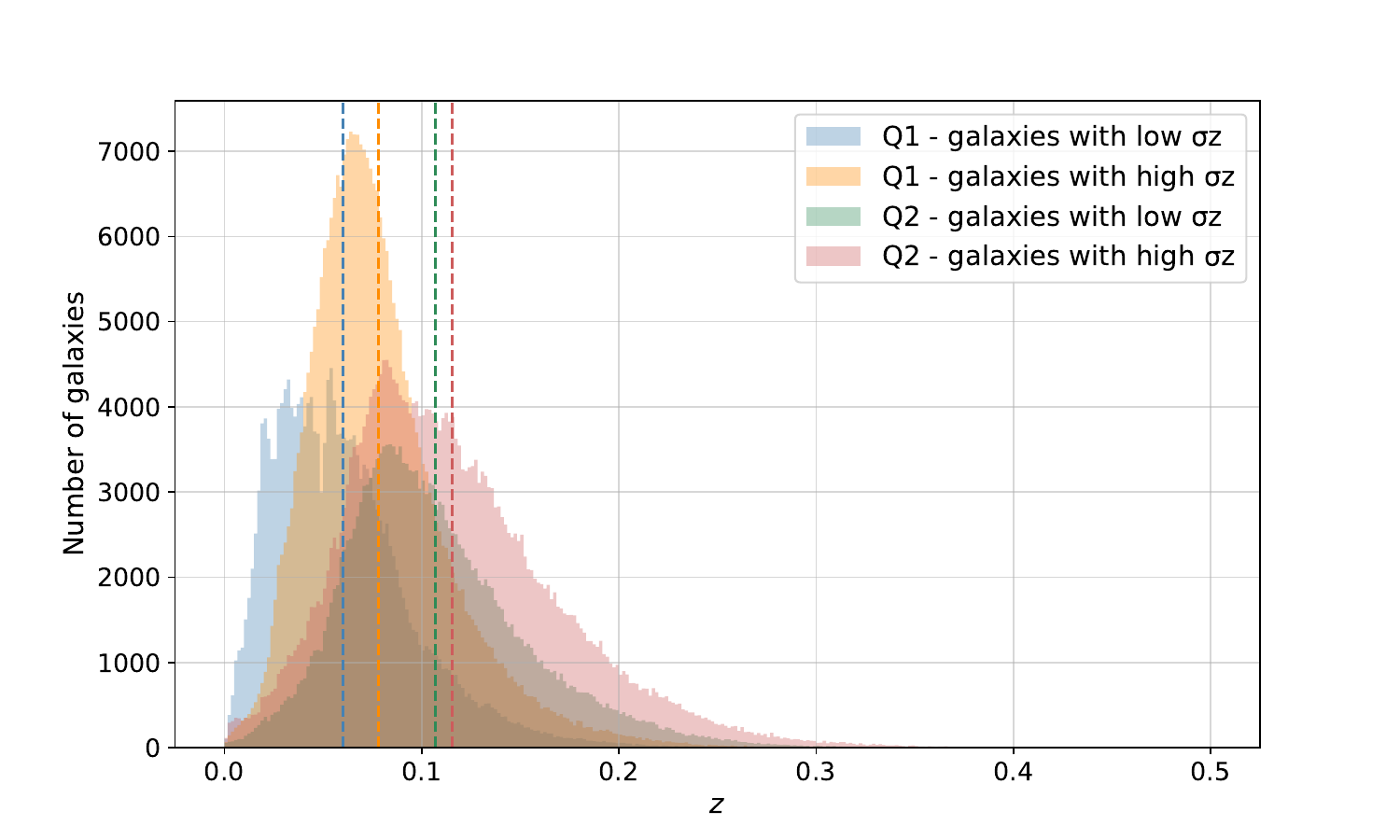}
    \caption{Distribution of redshift ($z$) for galaxies in the $K$-band subsample of GLADE+. 
    The sample is divided into bright (Q1) and faint (Q2) galaxies, each further split into 
    low-$\sigma_{z}$ and high-$\sigma_{z}$ populations. The histograms display the corresponding 
    redshift distributions, with dashed lines indicating the mean values.}
    \label{fig:redshift_subgroups}
\end{figure*}

Fig.~\ref{fig:redshift_subgroups} shows the histograms of the redshift distribution for the  four subgroups. The figure compares the  redshift distributions of the galaxies for these subgroups, with dashed lines indicating the mean values. The statistical properties of the redshift distributions for these four subgroups are summarized in Table~\ref{tab:redshift_subgroups}.

\begin{table*}[htbp]
\centering
\caption{Statistical summary of the redshift ($z$) distributions 
for the four subgroups defined by apparent magnitude and the bimodality in redshift error ($\sigma_{z}$). 
Confidence intervals (CI) correspond to the 68\% and 95\% percentiles.}
\label{tab:redshift_subgroups}
\begin{tabular}{lccccccc}
\hline
Group & $N$ & Mean $z$ & Median $z$ & 68\% CI  & 95\% CI &\\
\hline
Q1 -- low $\sigma_{z}$  & 198,608 & 0.06006 & 0.05450 & [0.02594, 0.09174] & [0.01061, 0.14829]\\
Q1 -- high $\sigma_{z}$ & 303,092 & 0.07819 & 0.07240 & [0.04610, 0.10932] & [0.02485, 0.16800]\\
Q2 -- low $\sigma_{z}$  & 67,968 & 0.10687 & 0.09938 & [0.06474, 0.15046] & [0.03229, 0.21289]\\
Q2 -- high $\sigma_{z}$ & 432,887 & 0.11561 & 0.10759 & [0.06535, 0.16800] & [0.02821, 0.24307]\\
\hline
\end{tabular}
\end{table*}


Overall, brighter galaxies lie at lower redshifts, while fainter galaxies extend systematically to higher $z$. The high-$\sigma_{z}$ subgroups show slightly higher redshifts than their low-$\sigma_{z}$ counterparts, as expected for photo-$z$ estimates at larger distances. Crucially, high $\sigma_{z}$ subgroups dominate the counts and exhibit broad, spread $z$ distributions, producing broad Gaussian peaks $p(z\mid\hat z)$ (see \ref{eq:2.4}) that dilute the constraining power of the in-catalogue term. Hence, increasing the $K$-band completeness percentile mainly incorporates high-$\sigma_{z}$, higher-$z$ galaxies and contributes little new information to the LOS prior, keeping its overall shape nearly unchanged.

\subsubsection{Redshift of GWTC-3 dark siren events and comparison with galaxy subgroups}
\label{sec:redshift_dark_sirens}

For the 46 GWTC-3 dark siren events, we read the posterior samples of the luminosity distance ($d_{L}$) from the released data \footnote{Posterior samples for the GWTC-3 events were obtained from the public
LIGO--Virgo--KAGRA parameter-estimation data release hosted on
\href{https://zenodo.org}{Zenodo}.}, and for each event, adopted the midpoint of the 68\% confidence interval as the distance indicator. We then converted $d_{L}$ to redshift using a flat $\Lambda$CDM cosmology with $H_{0}=70~\mathrm{km\,s^{-1}\,Mpc^{-1}}$ and $\Omega_{\mathrm{m}}=0.3$. The resulting summary statistics for the GWTC-3 dark sirens sample are provided in Table~\ref{tab:o3_redshift_summary}. 

In practice, the conversion from luminosity distance to redshift was performed using the \texttt{z\_at\_value} routine from the \texttt{astropy.cosmology} package\footnote{Astropy is available at \url{https://www.astropy.org/}}, which numerically inverts the $d_{L}(z)$ relation. Specifically, for each event we obtained
\begin{equation}
d_{L}(z) = (1+z)\,c \int_{0}^{z}\frac{dz'}{H(z')},
\end{equation}
with
\begin{equation}
H(z) = H_{0}\sqrt{\Omega_{\mathrm{m}}(1+z)^{3} + (1-\Omega_{\mathrm{m}})} ,
\end{equation}
corresponding to a flat $\Lambda$CDM model. The midpoint distance $d_{L,\mathrm{mid}}$ was then mapped to a redshift $z_{\mathrm{val}}$ using this relation.

\begin{figure*}[htbp]
    \centering
    \includegraphics[width=0.95\textwidth]{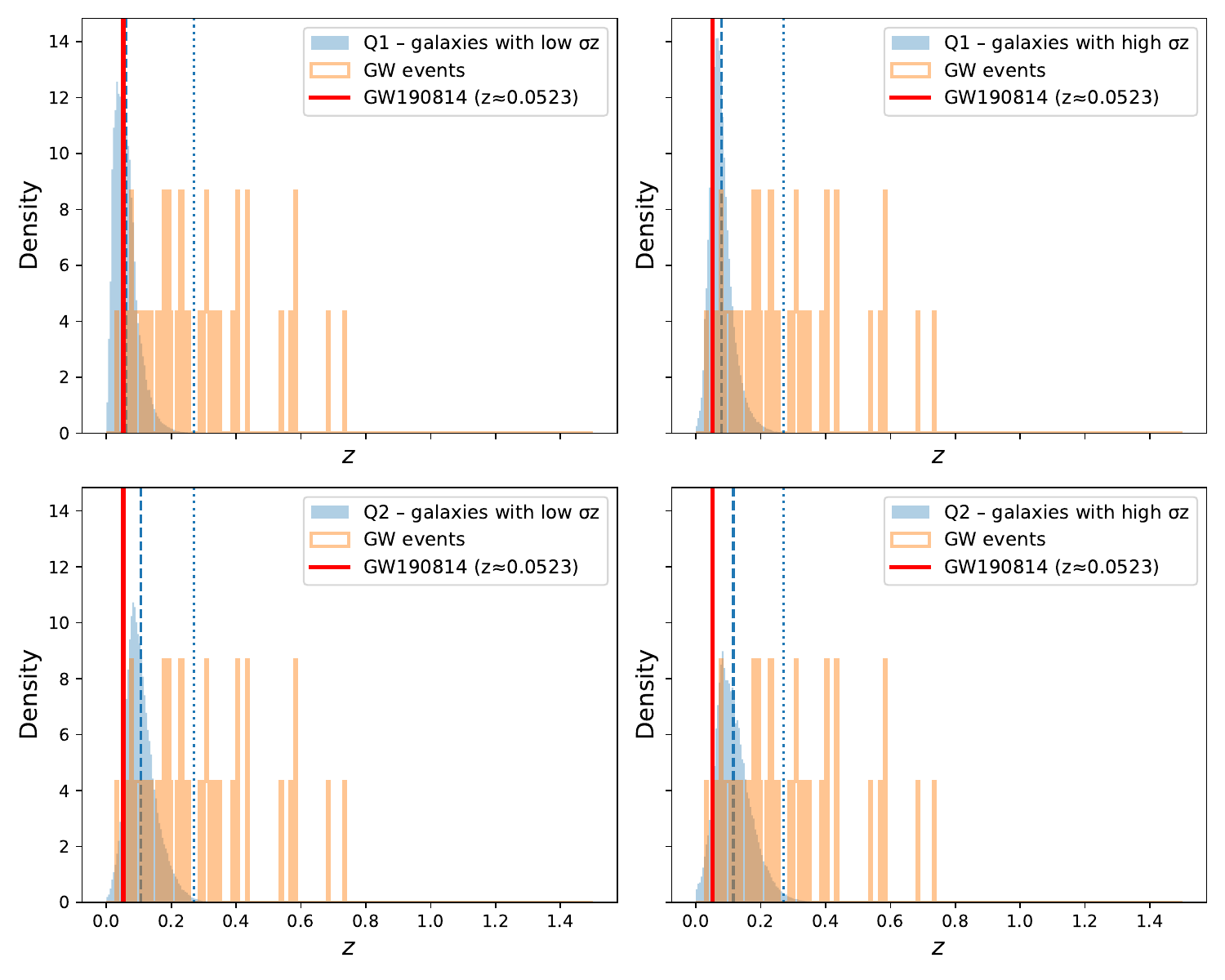}
    \caption{Redshift distributions for GLADE+ galaxies in four subsets (Q1/Q2 and low/high $\sigma_{z}$), with the GWTC-3 dark siren events overplotted. Shaded histograms show galaxy counts; the orange outline shows the GWTC-3 distribution, and vertical ticks mark individual event redshifts. The GW190814 event at $z = 0.0523$ is highlighted with a solid red line. Dashed and dotted lines indicate the means of the galaxy and event samples, respectively.}
    \label{fig:redshift_subgroups_gw}
\end{figure*}

To place these events in the context of the galaxy prior, we compared the GWTC-3 redshift distribution with the four galaxy subgroups defined in Sect.~\ref{sec:results_kband} (bright/faint and low/high $\sigma_{z}$; see Table~\ref{tab:redshift_subgroups} and Fig.~\ref{fig:redshift_subgroups}). Fig.~\ref{fig:redshift_subgroups_gw} overlays the distribution of the 46 dark siren events from GWTC-3 on the galaxies redshift histograms. GW190814 is highlighted to facilitate comparison in the nearby regime.

\begin{table*}[htbp]
\centering
\caption{Summary statistics for the GWTC-3 dark siren redshift sample. Confidence interval (CI) corresponds to the central 68\% interval.}
\label{tab:o3_redshift_summary}
\begin{tabular}{lccccc}
\hline
Sample & $N$ & Mean $z$ & Median $z$ & 68\% CI  \\
\hline
GWTC-3 dark sirens & 46 & 0.27028 & 0.23053 & [0.10064, 0.42641]\\
\hline
\end{tabular}
\end{table*}


From this comparison, GWTC-3 dark siren events typically lie at significantly higher redshifts than the bulk of galaxies with available $K$-band magnitudes. While the galaxy subgroups are concentrated below $z \sim 0.15$, the GWTC-3 sample extends to $z \gtrsim 0.4$, yielding a mean more than twice as high as that of the galaxy sets. In the context of the LOS prior, this mismatch matters because the $K$-band subcatalogue is not only sparse at these higher redshifts but also dominated by high-$\sigma_{z}$ galaxies; their large and systematic redshift uncertainties ($\sigma_{z}\!\sim\!0.015$) broaden $p(z \mid \hat z_k)$ and dilute the constraining power of the in-catalogue term. Consequently, increasing the $K$-band completeness percentile mainly adds more high-$\sigma_{z}$ galaxies at modest $z$, contributing little new information in the redshift range where most GWTC-3 events actually lie. As a result, the LOS prior relies heavily on its out-of-catalogue component for these events. Conversely, nearby events such as GW190814 fall within the well-sampled regime and provide a useful benchmark for assessing catalogue-completeness effects.

\subsection{Magnitude threshold maps ($m_{\rm th}$) across completeness percentiles}
\label{sec:results_mth}

\begin{figure}[htbp]
  \centering
  \includegraphics[width=0.45\textwidth]{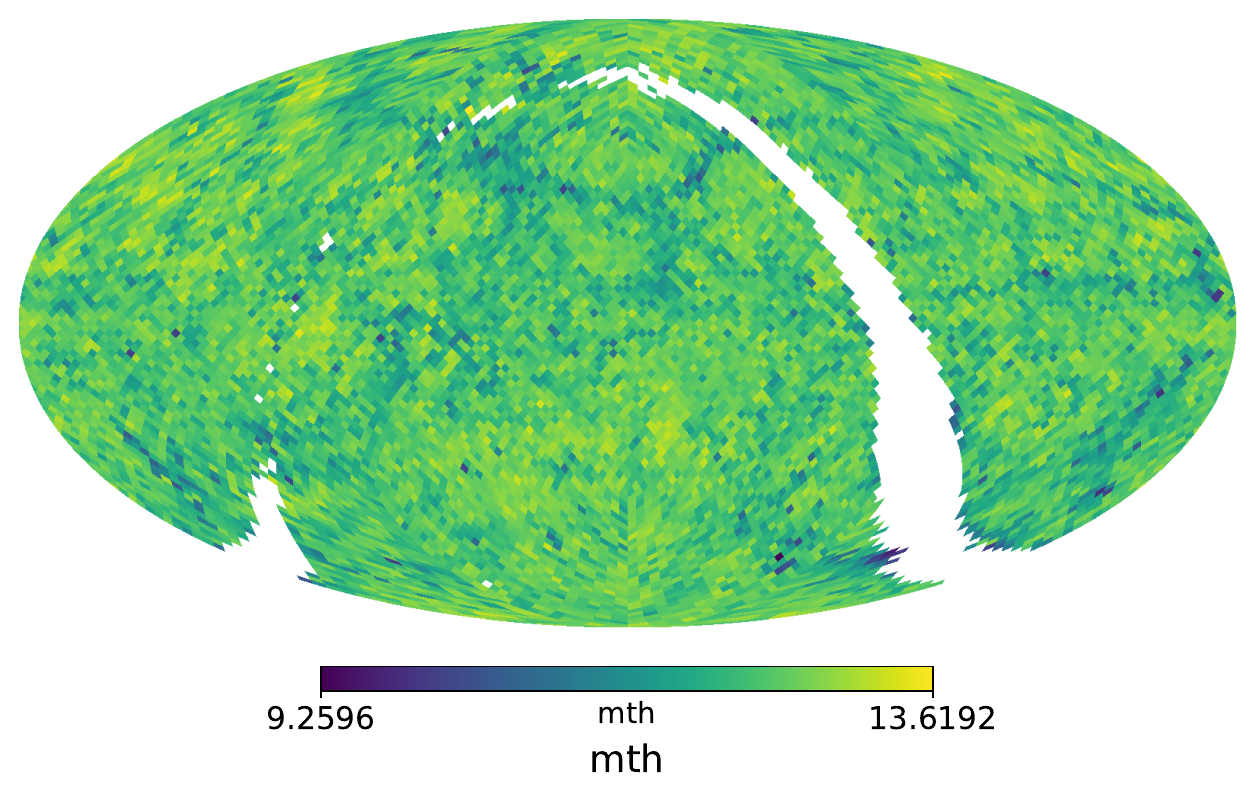}\\[0.5mm]
  \includegraphics[width=0.45\textwidth]{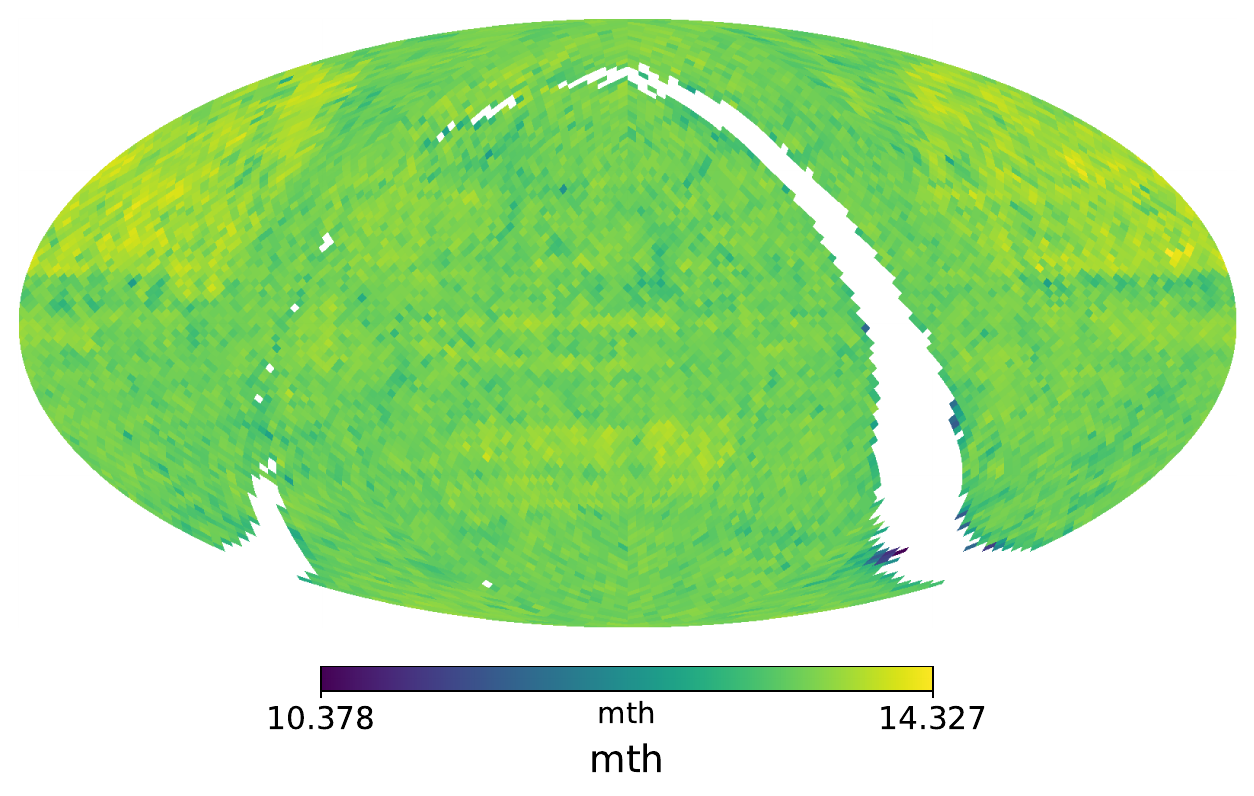}\\[2mm]
  \includegraphics[width=0.45\textwidth]{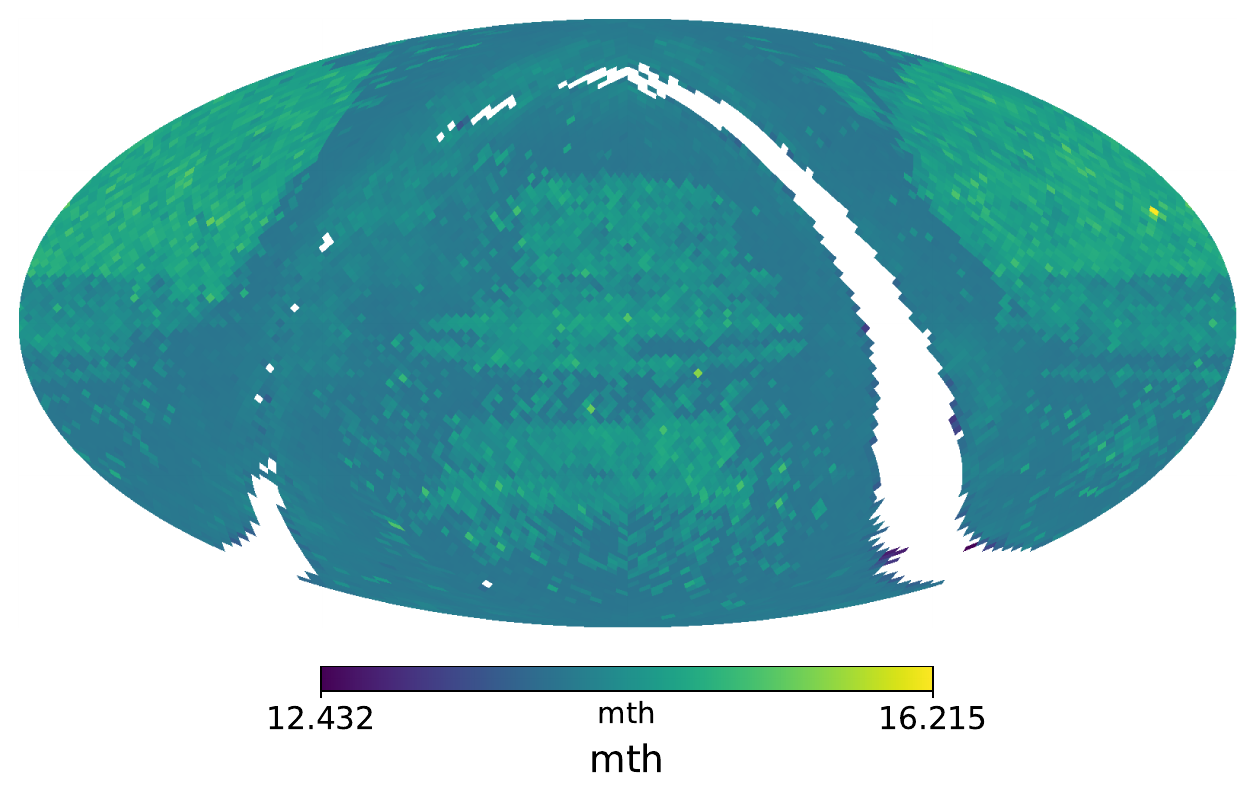}\\[2mm]
  \caption{Magnitude threshold (\textit{mth}) maps for completeness percentiles 10\% (top), 50\% (middle), and 100\% (bottom). Lower percentiles retain only the brightest galaxies, resulting in lower mth values and stronger incompleteness, while higher percentiles include fainter galaxies, increasing \textit{mth} across the sky. Color bars indicate the threshold magnitude in each pixel.}
  \label{fig:mth_maps_vertical}
\end{figure}

The magnitude threshold ($m_{\rm th}$) maps in Fig.~\ref{fig:mth_maps_vertical} illustrate how the limiting apparent magnitude per \textsc{healpix} \footnote{HEALPIX is available at \url{http://healpix.sf.net}} pixel changes when evaluating different completeness percentiles. As described in Section \ref{subsec:completeness_mth}, the completeness procedure modifies the $m_{\rm th}$ assigned to each pixel, which in turn determines the galaxies that contribute to the in-catalog and those that fall into the out-of-catalog component of the LOS construction. 

At the 10\% percentile (top panel) in Fig.~\ref{fig:mth_maps_vertical}, only the brightest galaxies are retained, leading to systematically lower $m_{\rm th}$ values and a stronger impact of incompleteness, particularly in low-density regions. At 50\% completeness (middle panel), the $m_{\rm th}$ increases across most of the sky, allowing fainter galaxies to be included and reducing incompleteness effects. Finally, at the 100\% percentile (bottom panel), the $m_{\rm th}$ reaches its maximum values, incorporating the faintest galaxies catalogued and yielding the deepest effective coverage. These variations highlight how the adopted completeness percentile directly controls the galaxy population entering the LOS prior, and consequently the balance between in-catalog and out-of-catalog contributions.

\subsection{Impact on LOS prior for an individual pixel}
\label{sec:results_pixel5000}

\begin{figure*}[htbp]
  \centering
  \includegraphics[width=0.90\textwidth]{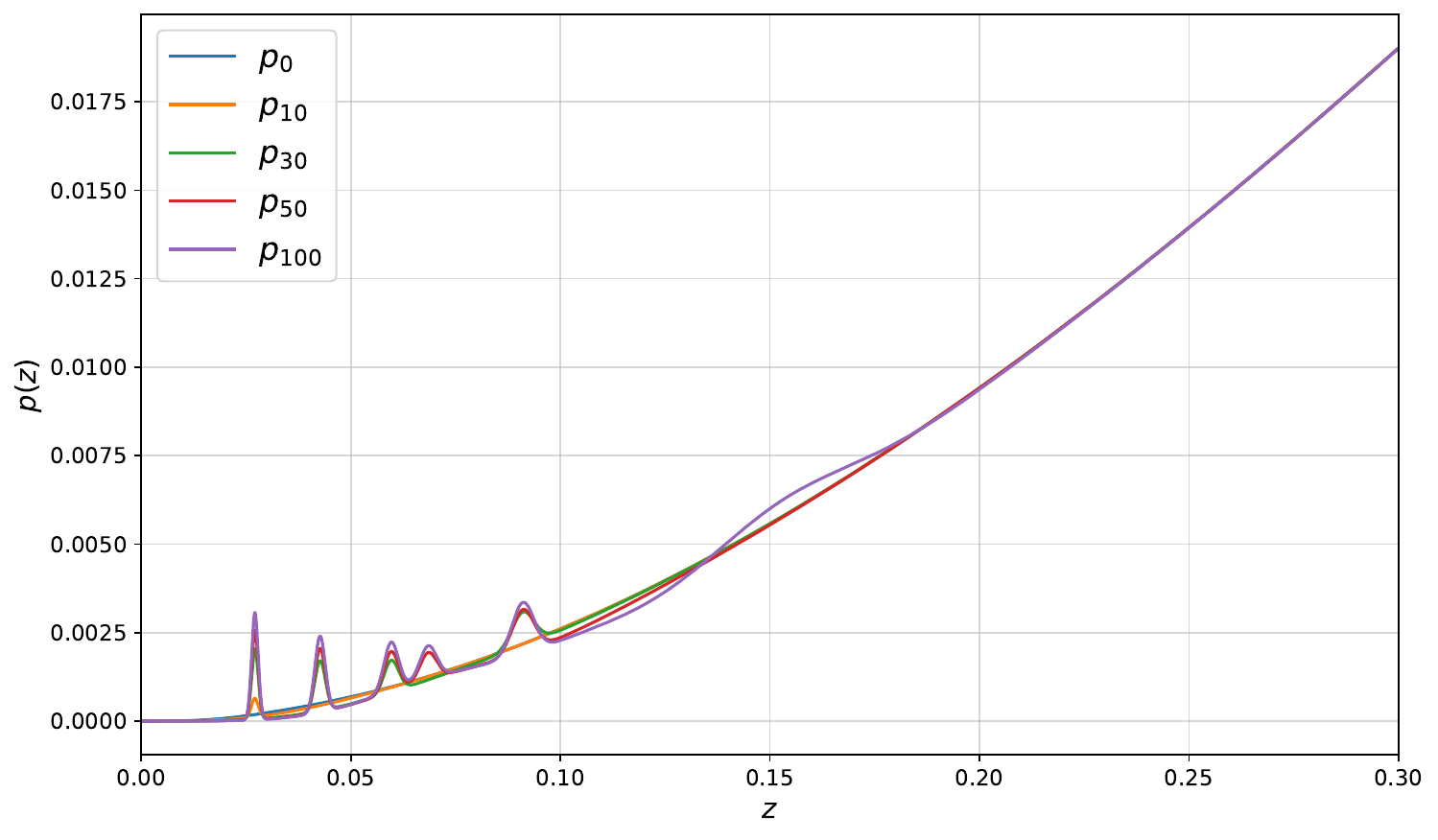}
  \caption{Comparison of the LOS prior $p(z)$ for pixel 17960 across different completeness percentiles, including the empty catalog case. Higher percentiles increase the number of galaxies with $m < m_{\mathrm{th}}$ and amplify the pixel’s contribution to the LOS prior.}
  \label{fig:los_pixel4500}
\end{figure*}

To further illustrate the local effect of catalog completeness on the LOS, we analyzed a representative \textsc{healpix} pixel (index 17960). Fig.~\ref{fig:los_pixel4500} compares the LOS $P(z)$ for this pixel at different completeness percentiles (0\%, 10\%, 30\%, 50\%, and 100\%). The corresponding values of $m_{\rm th}$ and the number of galaxies per percentile are summarized in Table~\ref{tab:pixel5000_stats}. The curves remain broadly consistent in shape, but their differences reflect the impact of including galaxies of different brightness and redshift quality. 

As discussed in Section \ref{sec:results_kband}, the galaxies that dominate in low percentiles (e.g., 10$\%$) belong mainly to Q1: the brightest sources with the smallest redshift errors $\sigma_{z}$. These provide important but relatively small contributions to the a priori distribution of the catalog. By increasing to the 50\% percentile, more of these bright galaxies are included, still characterized by low $\sigma_{z}$ values, so the prior gains information in a statistically reliable way. However, moving from the 50\% to the 100\% percentile incorporates the fainter galaxies from Q2. Although this step makes the catalog more complete, these galaxies present substantially larger $\sigma_{z}$ values. Their inclusion therefore adds little effective redshift information to the in-catalog term, despite increasing the number of sources. 

As emphasized in Section \ref{sec:redshift_dark_sirens}, the $K$-band catalog itself does not provide sufficient redshift coverage to reach the distances of the GWTC-3 dark siren events. This means that even when completeness is maximized by including the full Q2 population, the LOS priors remain unable to probe the redshift range relevant for these sources. This result is consistent with the idea that catalog completeness controls which galaxies contribute to the in-catalog term; the intrinsic depth of the $K$ band data imposes a fundamental limitation, leaving the out-of-catalog component dominant at the distances of interest.

\begin{table}[htbp]
\centering
\caption{Magnitude threshold $m_{\mathrm{th}}$, number of galaxies with $m < m_{\mathrm{th}}$ for pixel 17960 at different completeness percentiles.}
\label{tab:pixel5000_stats}
\begin{tabular}{cccc}
\hline
Percentile & $m_{\mathrm{th}}$ & $N(m < m_{\mathrm{th}})$ \\
\hline
10   & 11.71 & 9  \\
30   & 12.94 & 26 \\
50   & 13.34 & 42 \\
100  & 14.25 & 86 \\
\hline
\end{tabular}
\end{table}

\subsubsection{GW190814: sky-averaged ratio of LOS to empty LOS}\label{sec:gw190814_ratio}

\begin{figure*}[htbp]
    \centering
    \includegraphics[width=0.6\textwidth]{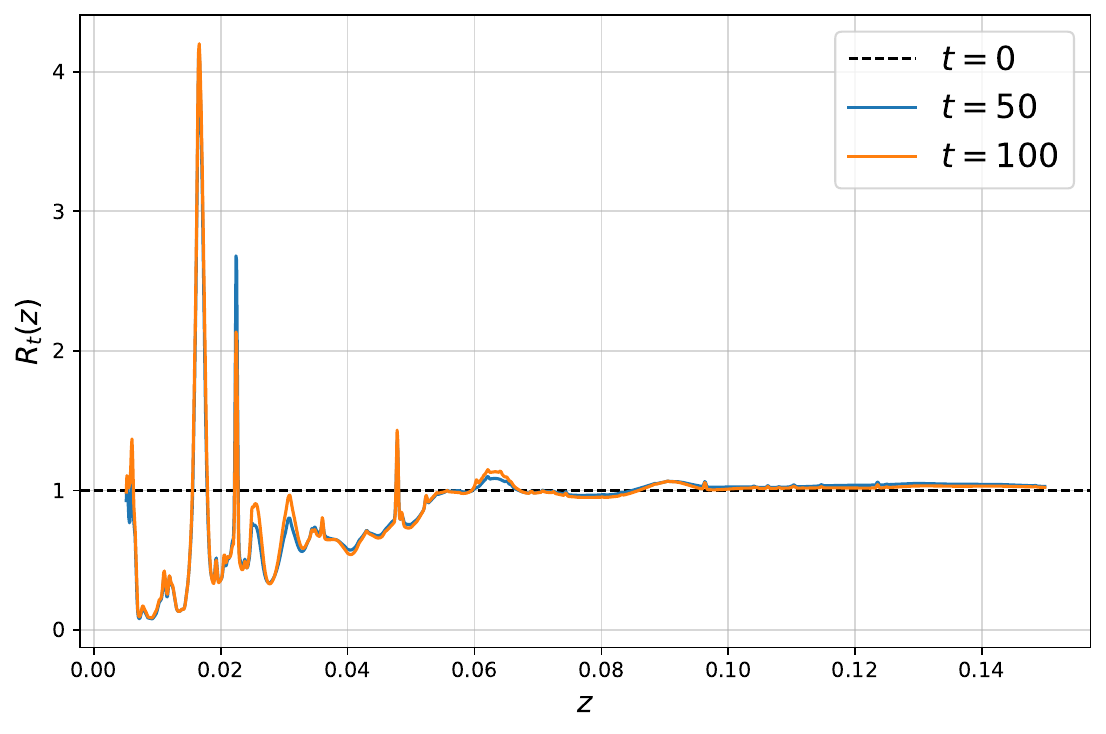}
    \includegraphics[width=0.6\textwidth]{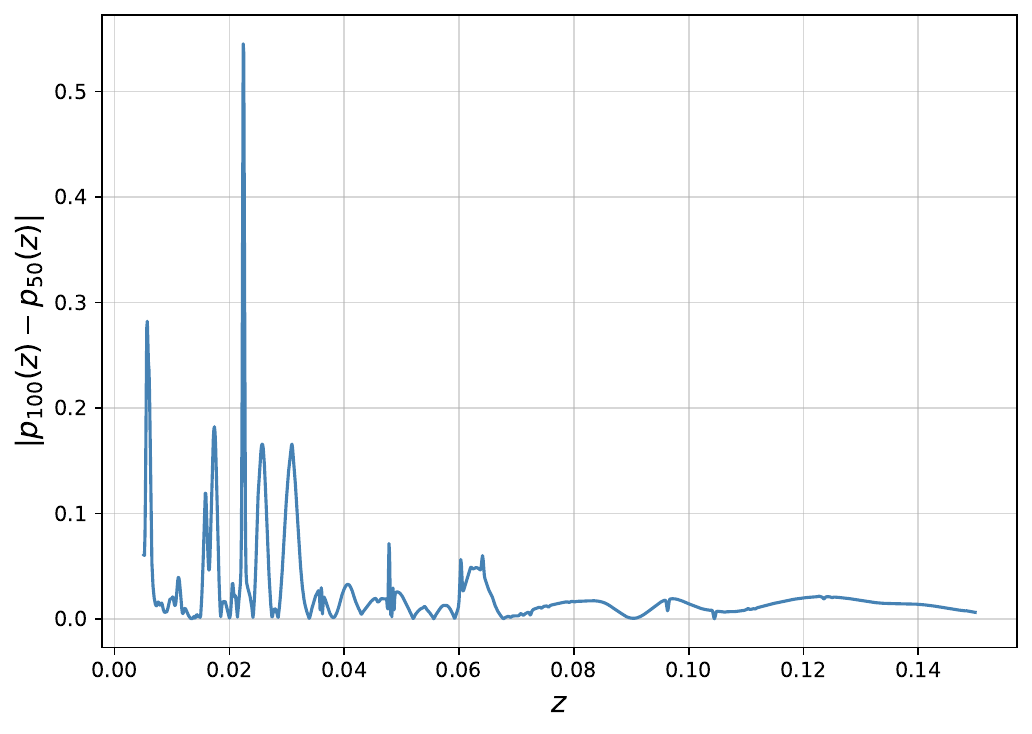}
    \includegraphics[width=0.6\textwidth]{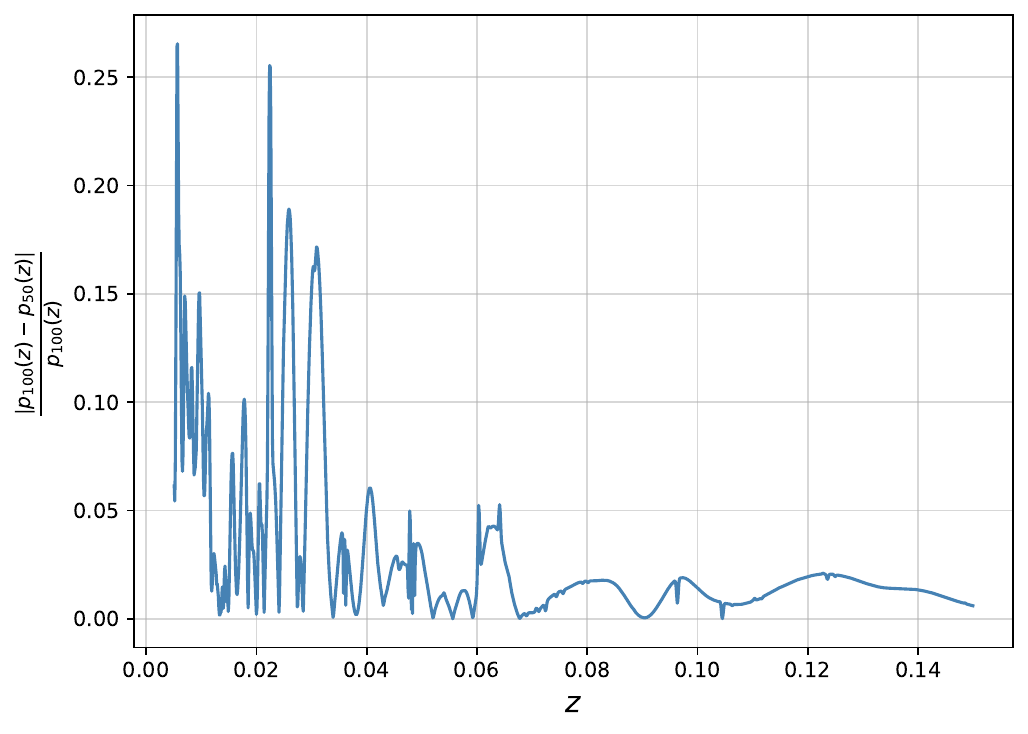}
    \caption{Comparison of LOS priors for the event localization of the event GW190814. 
The top panel shows $R_t(z)$ for different level of completeness. 
The middle panel shows the difference between $p_{50}(z)$ and $p_{100}(z)$. 
The bottom panel shows the relative difference between $p_{50}(z)$ and $p_{100}(z)$.}
    \label{fig:gw190814_fig9}
\end{figure*}

To quantify the informativeness of the LOS prior for a given event, we compute the ratio $R_{t}(z)$ between the single--event LOS prior and the sky--averaged LOS prior. 
Denoting the LOS prior in pixel $i$ as $p_{t}^{(i)}(z)$, we define the ratio $R_{t}(z)$ as

\begin{equation}
R_{t,e}(z) \;=\;
\frac{\displaystyle \sum_{i \in S} p_{t,e}^{(i)}(z)}
     {p_{0}(z)} \, ,
\label{eq:Rt}
\end{equation}
where $t$ denotes the apparent magnitude percentile threshold, $e$ stands for the event, $i$ is the pixel index, 
 $p_{0}(z)$ is the $0$ percentile prior, i.e. the empty catalogue prior, 
and $S$ denotes the set of pixels of the event sky localization.
Fig.~\ref{fig:gw190814_fig9} shows $R_{t}(z)$ for different values of $t$. 

At higher redshift ($z\!\gtrsim\!0.08$) different curves converge to $R_t(z)\!\simeq\!1$, since in that region the catalogue is incomplete. Since the fiducial redshift of GW190814  is within the redshift range where the catalog has good support for fainter galaxies, the effect on the LOS of including those galaxies is not negligible.

\subsection{Impact of catalog completeness on LOS priors}

\begin{figure*}[h]
  \centering
  \includegraphics[width=0.8\linewidth]{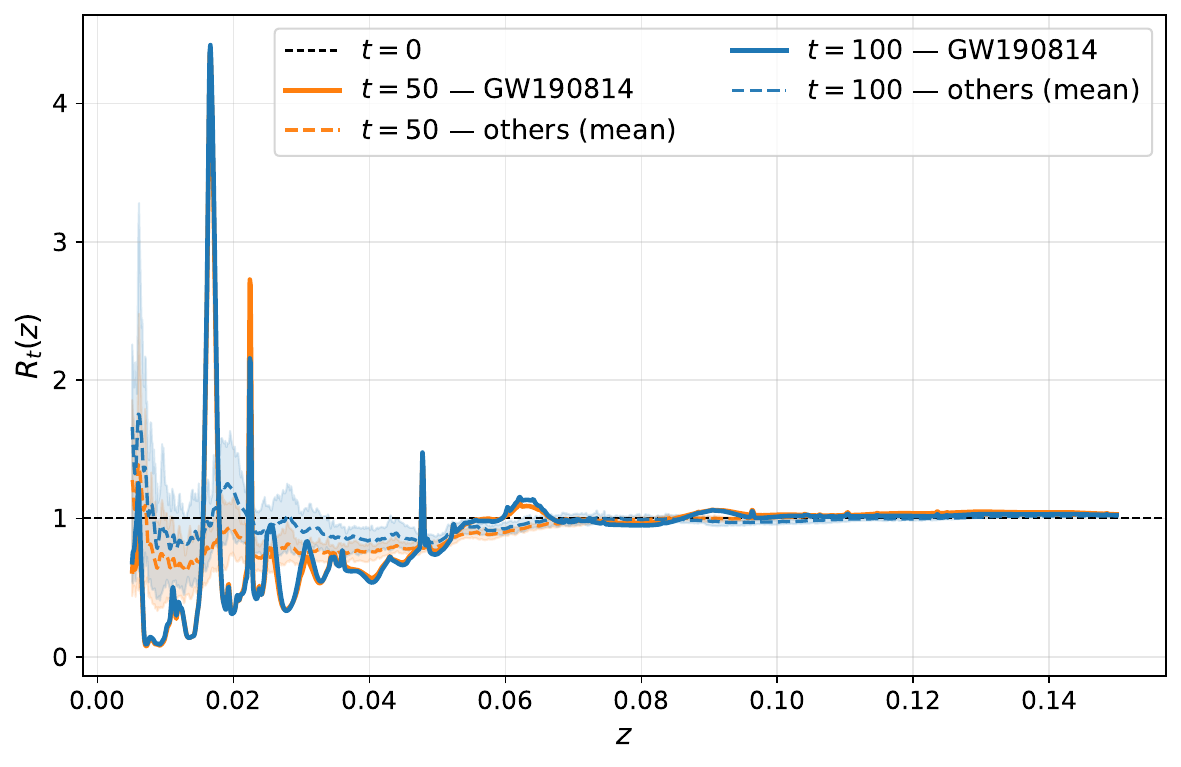}
  \caption{Comparison of $R_t(z)$ for the single event \texttt{GW190814} and the average of the other events, for which the shaded bands indicate the 68\% interval around the mean.}
  \label{fig:panelA}
\end{figure*}

In Fig.~\ref{fig:panelA}  we show the comparison between  the LOS prior of the event \texttt{GW190814} and the averages LOS prior of the other events.
For the average prior the catalog completeness has a negligible effect, while for \texttt{GW190814} there is a more noticeable effect. This is due to the good sky localization of this event, making the LOS prior more sensitive to the inclusion of fainter galaxies. In  Fig.~\ref{fig:panelA} the averaged LOS and its $68\%$ confidence bands are plotted as a function of redshift, showing the average impact of completeness. 


\subsubsection{Quantifying the difference between catalog and empty catalog LOS priors}

To quantify the difference between the catalog and empty catalog LOS redshift priors, respectively $p_t$ and $p_0$, we consider two statistical distance measures: the Kullback–Leibler (KL) divergence \cite{10.1214/aop/1176996454} and the Hellinger distance \cite{Hellinger+1909+210+271}.

The KL divergence is defined by 
\begin{equation}
\mathrm{KL}_{t} = \int p_t(z) 
\log\!\left[\frac{p_t(z)}{p_0(z)}\right] \, dz \,,
\end{equation}
while the Hellinger distance is given by
\begin{equation}
H_t =
\left[
\frac{1}{2}
\int
\left(
\sqrt{p_t(z)}-\sqrt{p_0(z)}
\right)^2
\, dz
\right]^{1/2}.
\end{equation}
For each gravitational-wave event, we also considered the sky localization area at 90\% credible level ($\Omega_{90}$, in deg$^{2}$), to investigate the relation between LOS prior informativeness and sky localization. The  results  are reported in Appendix \ref{sec:Indicators_appendix}, corresponding to the $50$ and $100$ percentile completeness thresholds.

\begin{figure*}[htbp]
    \centering
    \includegraphics[width=0.7\textwidth]{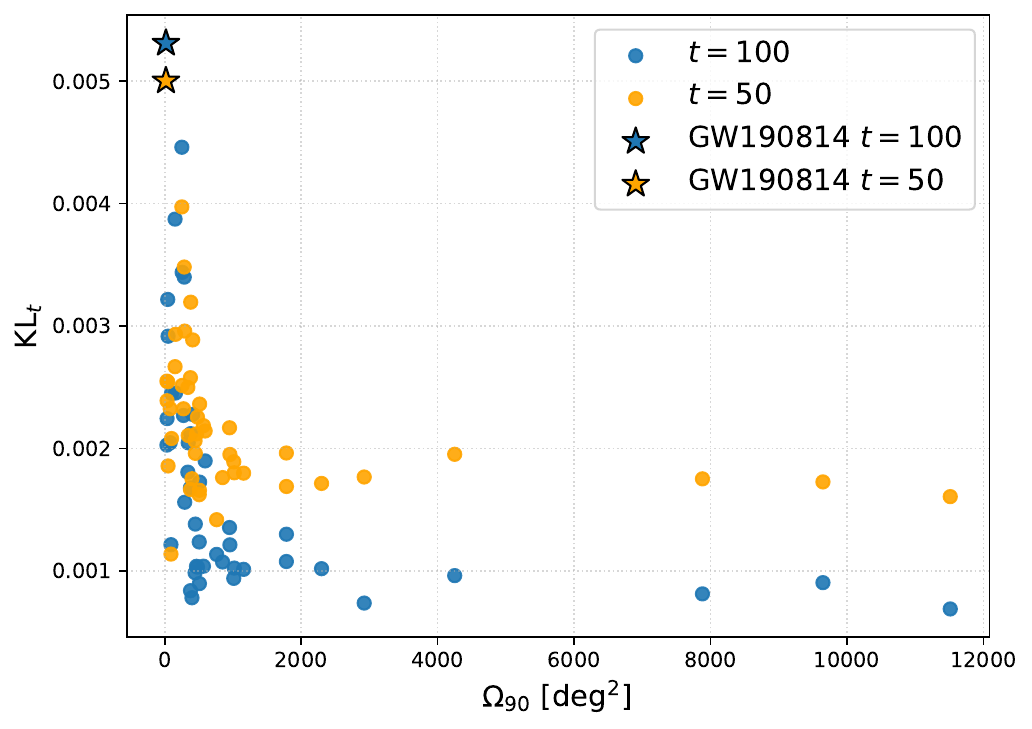}
    \includegraphics[width=0.7\textwidth]{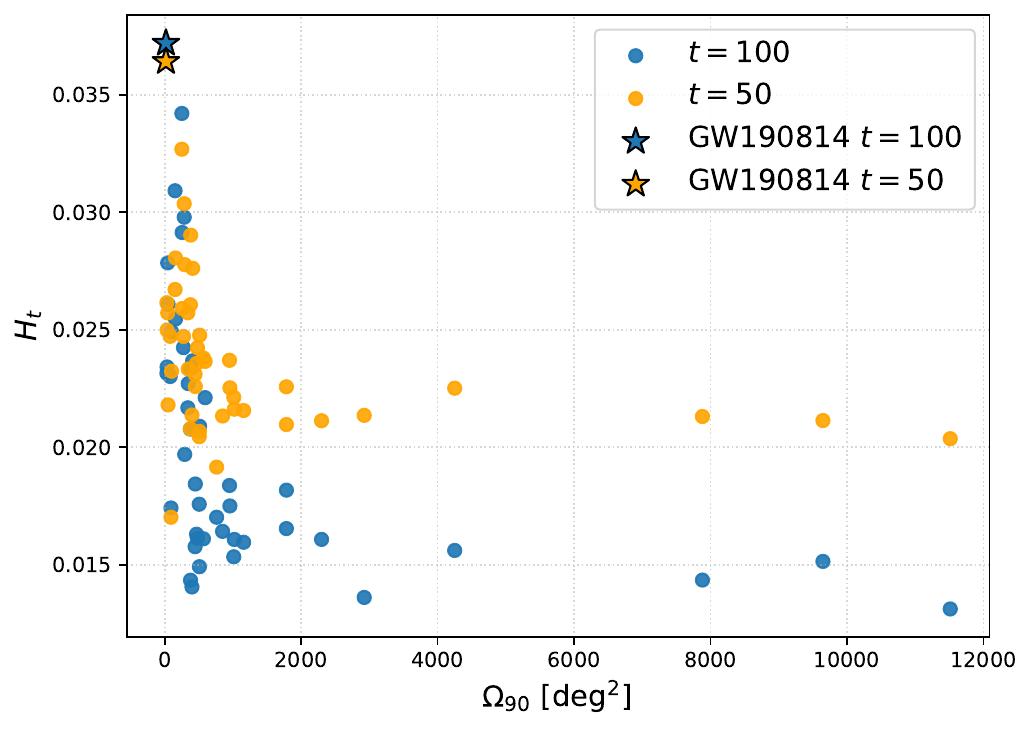}
    \caption{Statistical indicators of the LOS--empty differences as a function of the sky localization area $\Omega_{90}$. Each panel shows the results for the 50th (orange) and 100th (blue) percentiles. Each marker represents an GWTC-3 dark siren event.}
    \label{fig:indicators_vs_A90}
\end{figure*}

In Fig.~\ref{fig:indicators_vs_A90} we plot the informativeness indicator $H_t$ for the different events versus their $\Omega_{90}$. 

For GW190814 (denoted with a star in Fig.~\ref{fig:indicators_vs_A90}) the LOS prior informativeness is higher than for other events, since it has a small sky localization area and it is located inside the catalog redshift range. 

\subsection{Impact of Redshift Uncertainty of dimmer galaxies}
\label{sec:redshift_uncertainty_impact}
As shown in the previous sections, the results obtained including dimmer galaxies in the LOS calculations are very similar to those obtained using only bright galaxies.
We have shown that dimmer galaxies tend to have larger redshift errors, and this could be the cause of the negligible effect of their inclusion in the analysis. To test this hypothesis we create a galaxy catalog with the same objects present in Q2, but we change their redshift errors, sampling them from the distribution of the Q1 redshift errors.

\begin{figure*}[t]
    \centering
    \includegraphics[width=0.7\textwidth]{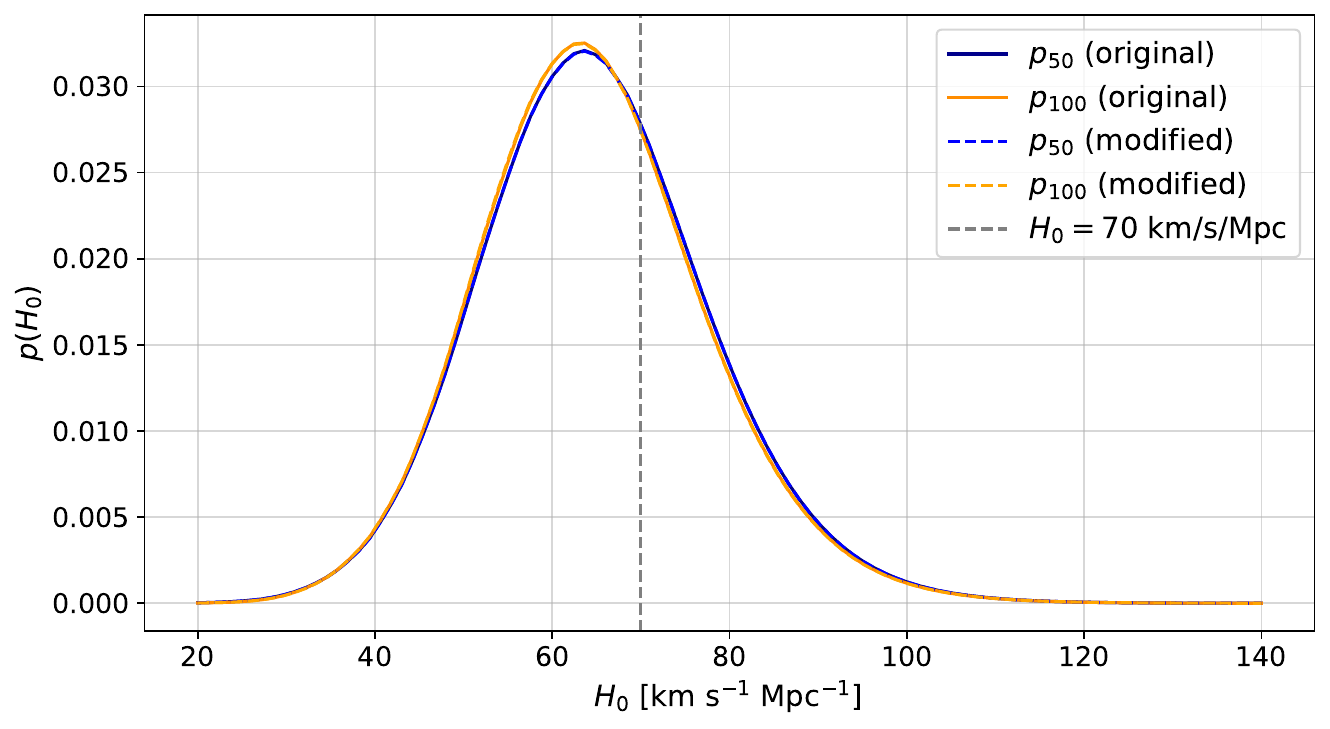}
    \caption{Joint $H_0$ posterior from 42 BBHs. Solid lines correspond to LOS redshift priors constructed using the original redshift uncertainty distributions, while dashed lines represent LOS priors built from the redshift error distribution modified as described in Sec.~\ref{sec:redshift_uncertainty_impact}. Results are shown for both $p_{50}$ and $p_{100}$ completeness levels.
    }
    \label{fig:O3_joint_modified_test}
\end{figure*}

\begin{figure*}[htbp]
    \centering
    \includegraphics[width=0.6\textwidth]{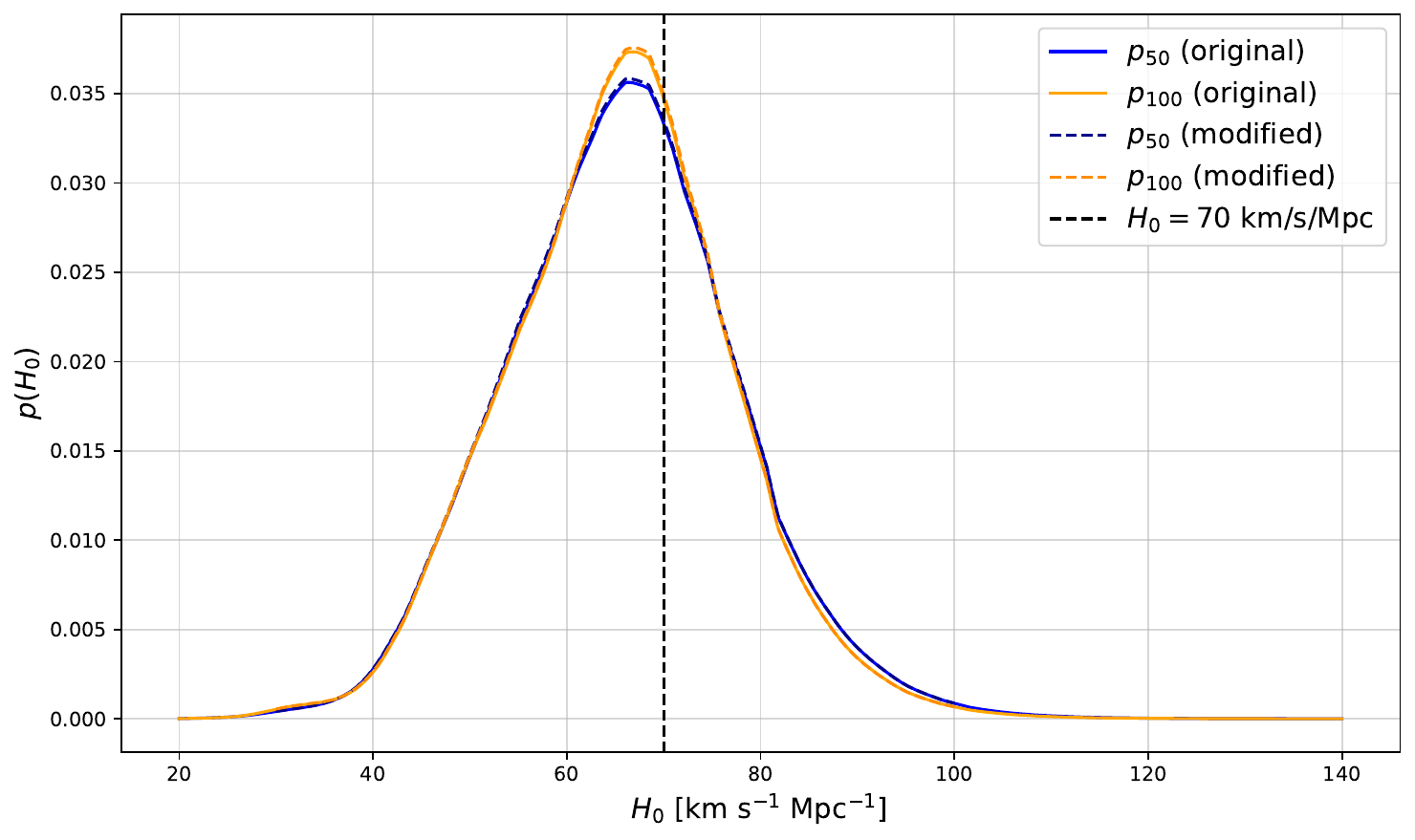}
    \caption{Combined $H_0$ posterior for 46 events (42 BBHs + 3 NSBHs + GW190425 ) for different value of the completeness percentile. 
    Solid lines correspond to LOS priors constructed using the original redshift uncertainty 
    distribution, while dashed lines represent the  redshift uncertainty modified as described in Sec.~\ref{sec:redshift_uncertainty_impact}. 
    Results are shown for both $p_{50}$ and $p_{100}$ completeness levels.}
    \label{fig:O3_combined_modified_test}
\end{figure*}

\begin{figure*}[t]
    \centering
    \includegraphics[width=0.7\textwidth]{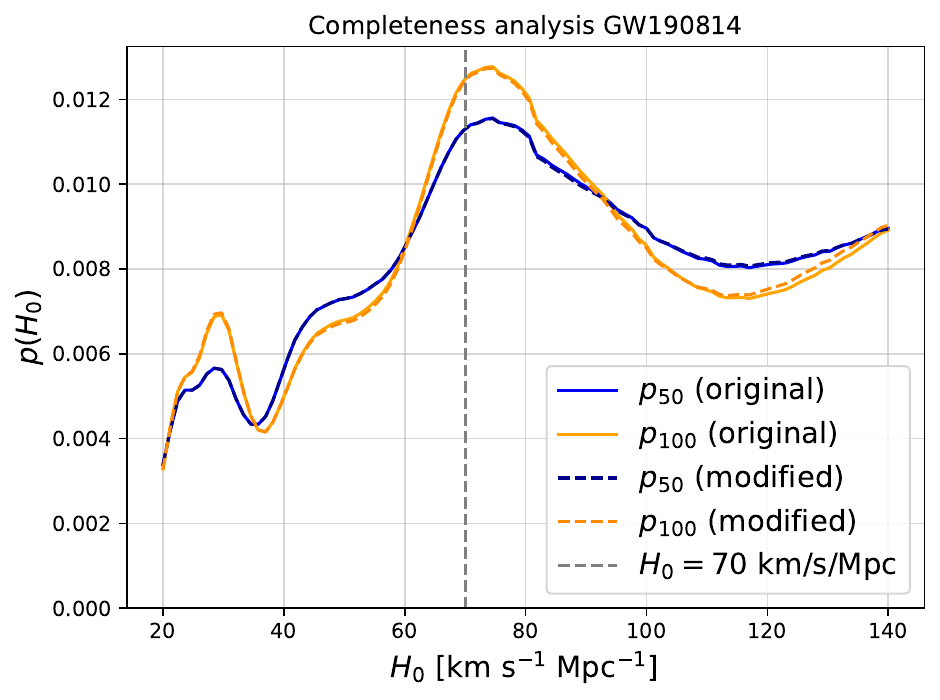}
    \caption{Posterior distributions of $H_{0}$ for GW190814 using LOS redshift priors constructed from the original 
    and modified redshift uncertainty distributions. Solid lines correspond to the original 
    LOS construction, while dashed lines represent the modified case. Results are shown 
    for $p_{50}$ and $p_{100}$ completeness levels.
    }
    \label{fig:GW190814_modified_test}
\end{figure*}

\begin{figure*}[t]
    \centering
    \includegraphics[width=0.8\textwidth]{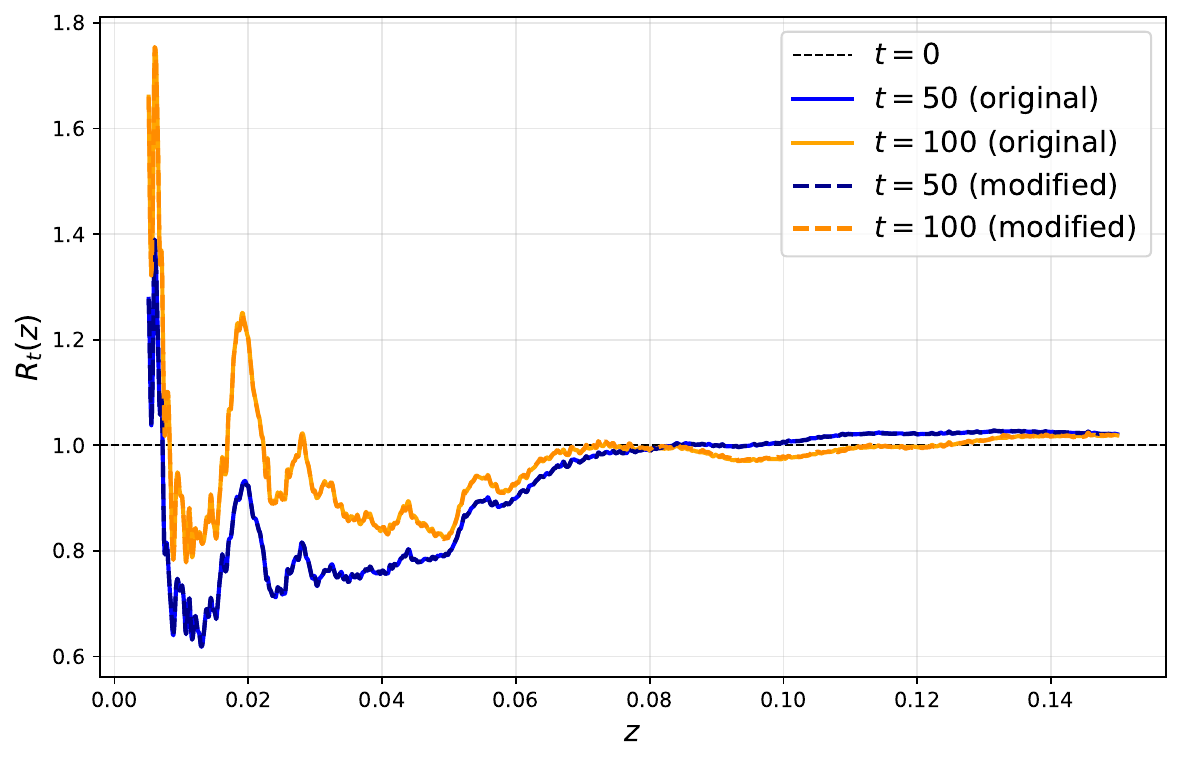}
    \caption{The quantity $R_{t,e}(z)$ defined in eq.(\ref{eq:Rt}), averaged over all the GWTC-3 dark sirens events, except  GW190814, is plotted as a function of the redshift for $t = 50$ and $t = 100$. Solid lines correspond to LOS priors constructed using the original redshift uncertainty distribution, while dashed lines represent the modified case.}
    \label{fig:Rt_modified_test}
\end{figure*}

As can be seen in Figs.~\ref{fig:O3_joint_modified_test}-\ref{fig:Rt_modified_test} and Tables~\ref{tab:O3_joint_modified_test}--\ref{tab:GW190814_modified_test}, the results obtained including the observed and modified Q2 data are very similar, from which we can conclude that the larger redshift errors of dimer galaxies  is not the main factor determining their negligible impact on the estimation of cosmological parameters.

\begin{table*}[t]
\centering
\caption{Joint results on $H_0$ from the GWTC-3 sample of 42 BBH events using original and modified LOS redshift priors at different completeness levels.}
\label{tab:O3_joint_modified_test}
\begin{tabular}{lcccccc}
\hline
Percentile & $H_0$ [km s$^{-1}$ Mpc$^{-1}$] & 68\% CI & 95\% CI & Rel. Err. (68\%) & Rel. Err. (95\%) \\
\hline
50 (original)  & $63.64^{+13.24}_{-11.96}$ & (51.68, 76.88) & (40.31, 91.79) & 39.60\% & 80.91\% \\
50 (modified)  & $63.64^{+13.25}_{-11.96}$ & (51.68, 76.89) & (40.32, 91.79) & 39.61\% & 80.87\% \\
100 (original) & $63.64^{+12.76}_{-12.15}$ & (51.49, 76.40) & (40.33, 91.36) & 39.14\% & 80.18\% \\
100 (modified) & $63.64^{+12.72}_{-12.17}$ & (51.47, 76.36) & (40.34, 91.33) & 39.11\% & 80.12\% \\
\hline
\end{tabular}
\end{table*}

\begin{table*}[t]
\centering
\caption{Combined single–event constraints on $H_0$ for the GWTC-3 sample (46 events, $n_{\rm side}=64$), using original and modified LOS redshift priors at different completeness levels.}
\label{tab:O3_combined_modified_test}
\begin{tabular}{lccccc}
\hline
Percentile & $H_0$ [km s$^{-1}$ Mpc$^{-1}$] & 68\% CI & 95\% CI & Rel. Err. (68\%) & Rel. Err. (95\%) \\
\hline
50 (original)  & $66.27^{+10.77}_{-12.51}$ & (53.76, 77.04) & (43.02, 90.16) & 35.13\% & 71.14\% \\
50 (modified)  & $66.27^{+10.75}_{-12.51}$ & (53.77, 77.03) & (43.01, 90.26) & 35.15\% & 71.21\% \\
100 (original) & $66.79^{+9.61}_{-13.00}$  & (53.79, 76.40) & (43.00, 88.83) & 33.85\% & 68.62\% \\
100 (modified) & $66.79^{+9.34}_{-13.09}$  & (53.84, 76.27) & (43.02, 88.85) & 33.80\% & 68.73\% \\
\hline
\end{tabular}
\end{table*}

\begin{table*}[t]
\centering
\caption{Constraints on $H_0$ from GW190814 using original and modified LOS redshift priors at different completeness levels.}
\label{tab:GW190814_modified_test}
\begin{tabular}{lccccc}
\hline
Percentile & $H_0$ [km s$^{-1}$ Mpc$^{-1}$] & 68\% CI & 95\% CI & Rel. Err. (68\%) & Rel. Err. (95\%) \\
\hline
50 (original)  & $74.55^{+46.66}_{-26.38}$ & $(48.17, 121.21)$ & $(24.34, 137.20)$ & 97.98\% & 151.39\%  \\
50 (modified)  & $74.55^{+46.68}_{-26.39}$ & $(48.16, 121.23)$ & $(24.35, 137.20)$ & 98.02\%  & 151.38\%  \\
100 (original) & $74.55^{+46.68}_{-26.73}$ & $(47.82, 120.23)$ & $(24.18, 137.18)$ & 97.26\%  & 151.58\%  \\
100 (modified) & $74.55^{+46.17}_{-26.74}$ & $(47.81, 120.72)$ & $(24.18, 137.22)$ & 97.82\%  & 151.63\%  \\
\hline
\end{tabular}
\end{table*}

\section{Conclusions}
We have investigated the effects of including fainter galaxies in the calculation of the LOS prior used to analyze the GWTC-3 events. For the multi-events analysis we did  not find any significant effect due to including fainter galaxies.
For single event analyses the effects are also mostly negligible, except for  GW190814.

The main outcome of our study is that the most important factors are a small sky localization area, and an overlap between the fiducial GW event redshift (derived from their distance under a fiducial cosmology) and the galaxy catalog redshift range.

In Section~\ref{sec:redshift_uncertainty_impact} we have shown that, despite fainter galaxies having larger redshift uncertainties compared to brighter galaxies, this is not the main factor determining their negligible impact on the estimation of cosmological parameters. The differences between the original and modified LOS constructions are minimal, indicating that redshift accuracy alone does not explain the completeness-dependent behavior of $H_0$.

For GW190814 the effects are more relevant, because it is has a good angular localization ($\sim 18\, {\rm degrees}^2$) and is at a distance with good overlap with fainter galaxies catalog coverage.
To check quantitatively what is the information gained by adding fainter galaxies we have computed two measures of the difference between the empty and the catalog LOS priors, confirming that for  GW190814 the inclusion of fainter galaxies produces a significant difference in the LOS, reducing the relative error of the $H_0$ estimation.

In the future it will be interesting to extend our analysis to other GW events, such as those of GWTC-4 \citep{GWTC4}, using deeper galaxies catalogs, to check what are the effects of including fainter galaxies on the informativeness of the LOS and on the estimation of cosmological parameters.

\section{ACKNOWLEDGMENTS}
This material is based upon work supported by NSF's LIGO Laboratory,
which is a major facility fully funded by the National Science Foundation.

\clearpage

\bibliography{src/references.bib}{}
\bibliographystyle{aasjournal}

\appendix
\renewcommand{\thesection}{\Alph{section}}
\renewcommand{\thesubsection}{\thesection.\arabic{subsection}}

\section{Results of the analysis with \texorpdfstring{$n_{\mathrm{side}}=32$}{nside = 32}}
\label{sec:Analysis_LOS_nside_32}

\begin{figure*}[htbp]
  \centering
  \includegraphics[width=0.8\textwidth]{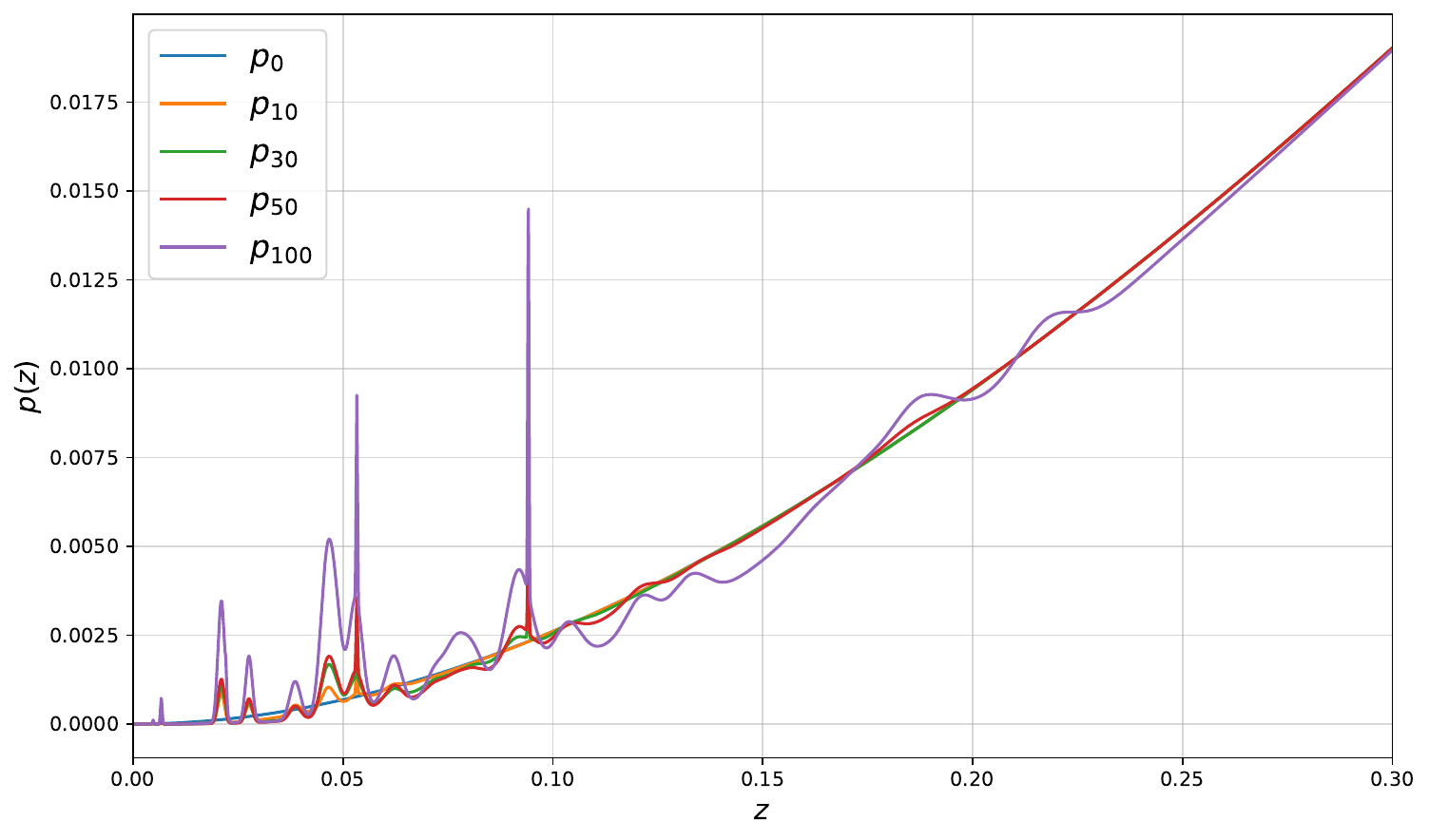}
  \caption{Comparison of the LOS prior $p(z)$ for pixel 4500 across different completeness percentiles, including the empty catalog case. This pixel corresponds the same sky position of the pixel 17960 with nside 64.}
  \label{Fig:LOS_nside32}
\end{figure*}

\begin{figure}[htbp]
    \centering
    \includegraphics[width=0.6\textwidth]{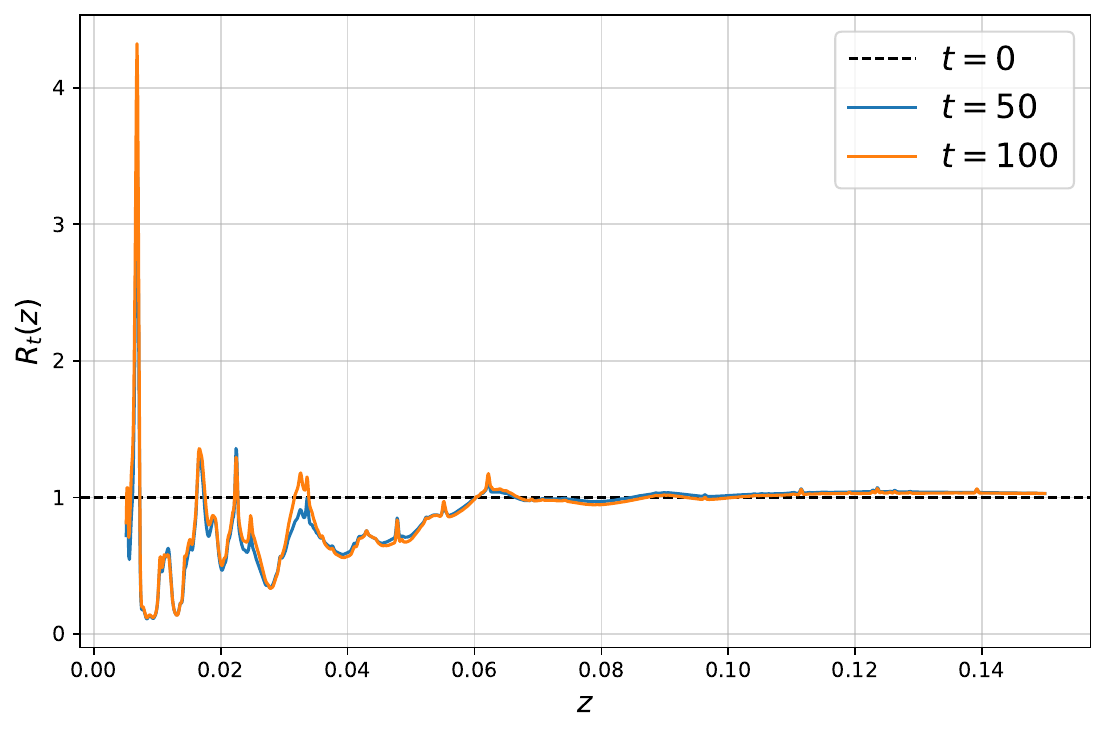}
    \includegraphics[width=0.6\textwidth]{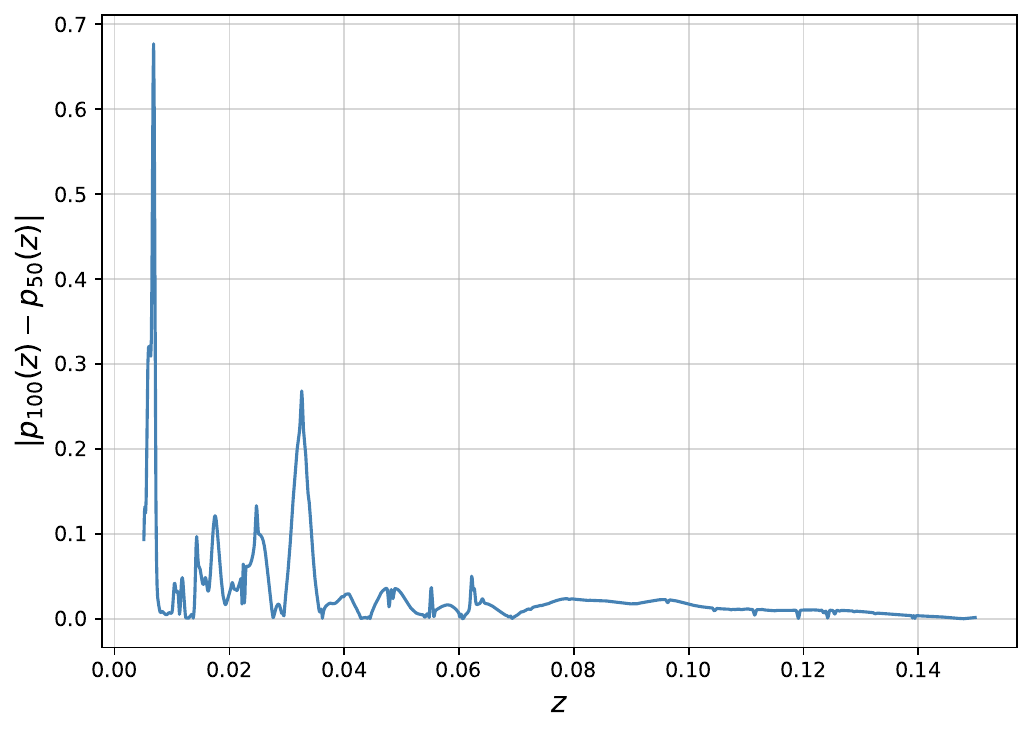}
    \includegraphics[width=0.6\textwidth]{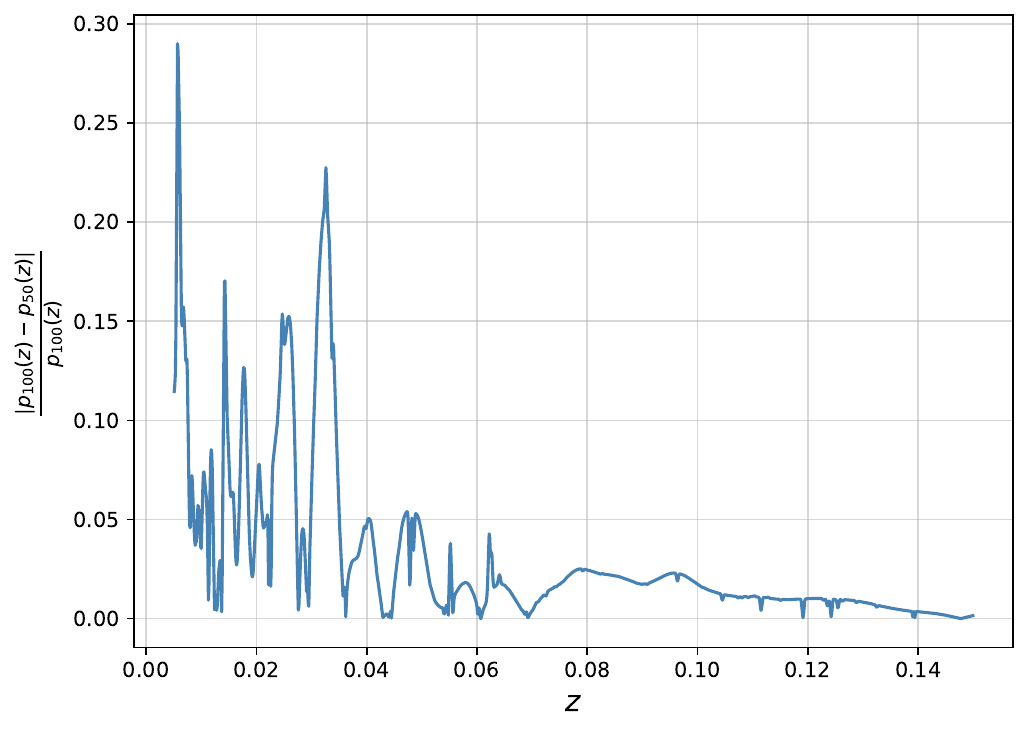}
    \caption{Comparison of LOS priors for the event localization of the event GW190814. 
The top panel shows $R_t(z)$ for different level of completeness. 
The middle panel shows the difference between $p_{50}(z)$ and $p_{100}(z)$. 
The bottom panel shows the relative difference between $p_{50}(z)$ and $p_{100}(z)$.}
\label{Fig:RT_nside32}
\end{figure}

\begin{figure*}[htbp]
  \centering
  \includegraphics[width=0.8\linewidth]{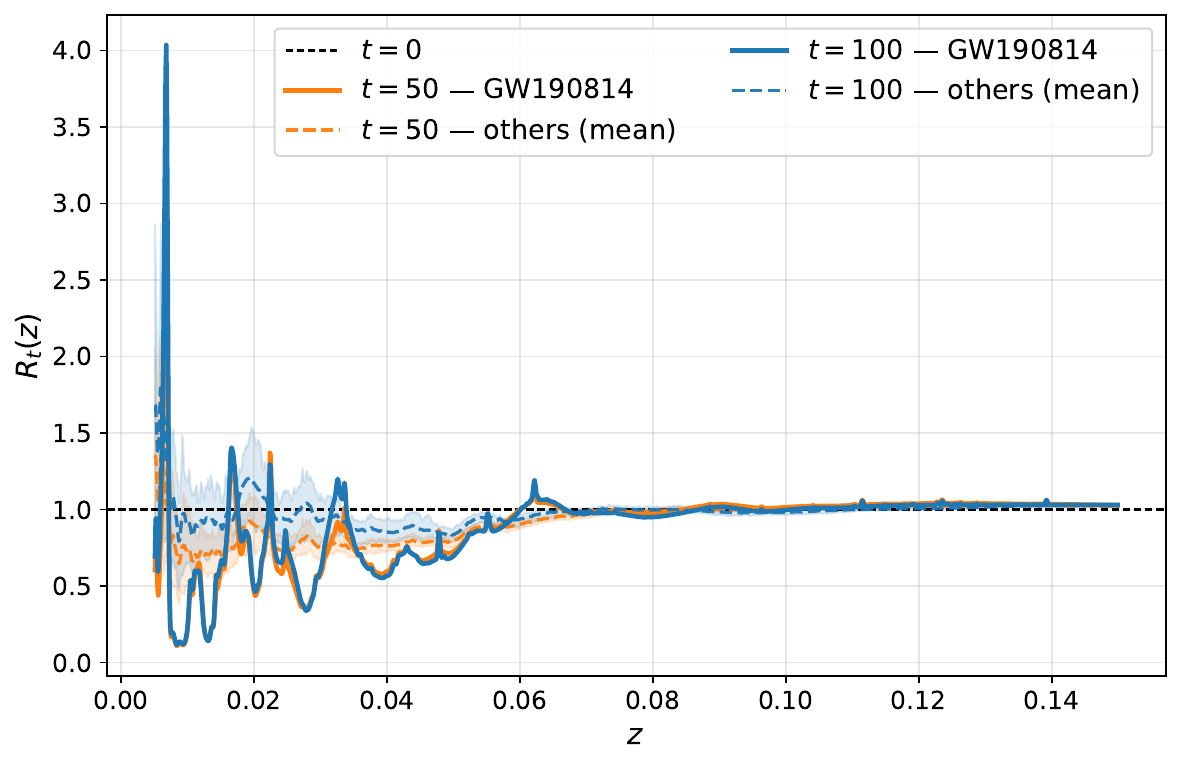}
  \caption{Comparison of $R_t(z)$ for the single event \texttt{GW190814} and the average of the other events, where for the latter  the shaded bands indicate the 68\% interval with respect to the mean.}
\label{Fig:single_RT_nside32}
\end{figure*}

\begin{figure*}[htbp]
    \centering
    \includegraphics[width=0.8\textwidth]{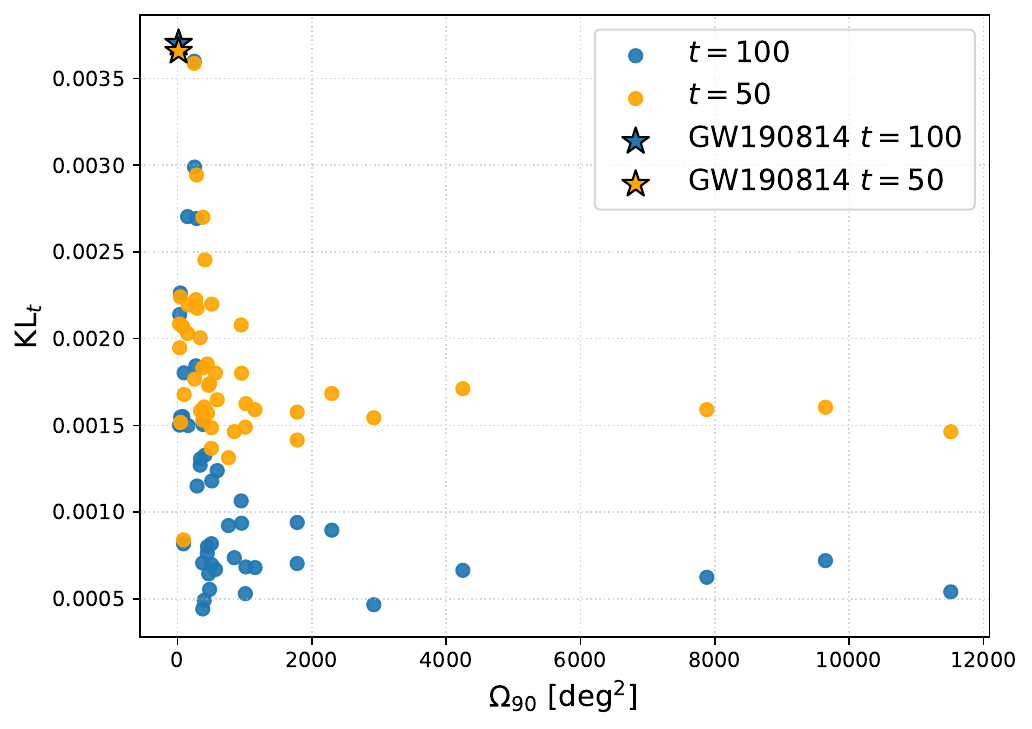}
    \includegraphics[width=0.8\textwidth]{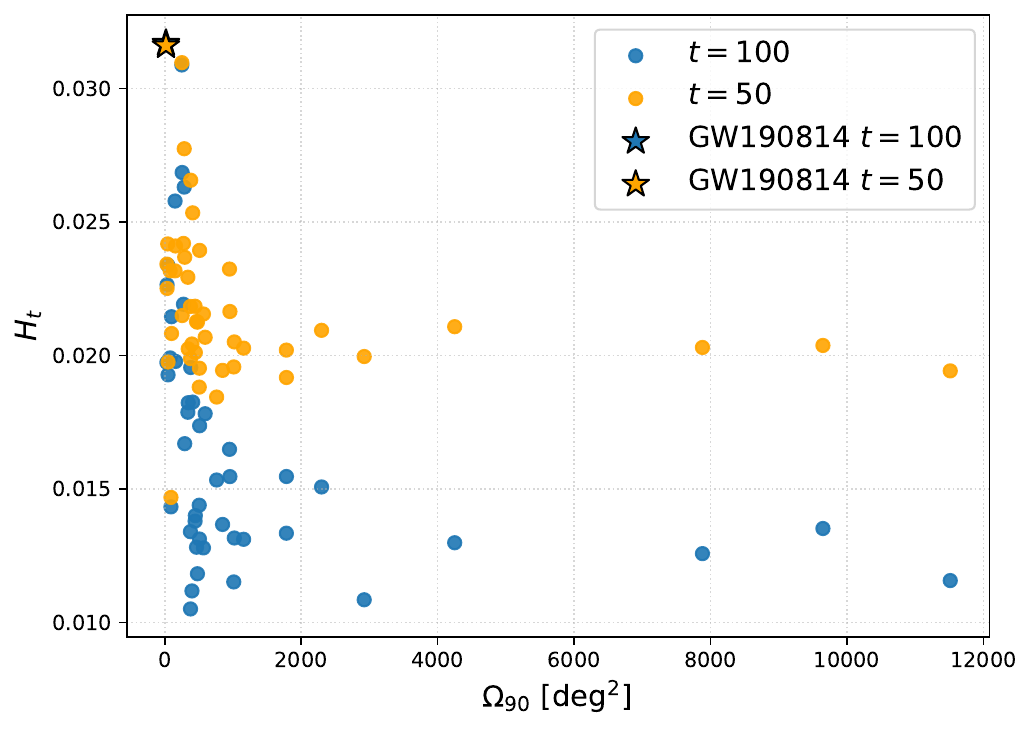}
    \caption{Statistical indicators of the LOS--empty differences as a function of the sky localization area $\Omega_{90}$. Each panel shows the results for the 50th (orange) and 100th (blue) percentiles. Each dot represents an GWTC-3 dark siren event}
    \label{Fig:Indicators_nside32}
\end{figure*}

In this appendix we report the results of the analysis with \texorpdfstring{$n_{\mathrm{side}}=32$}{nside = 32}, to check the impact of the LOS resolution.
Figs.~\ref{Fig:LOS_nside32}-\ref{Fig:Indicators_nside32} show the LOS $p(z)$ for different completeness percentiles.

\begin{figure}[h]
  \centering
  \includegraphics[width=0.85\textwidth]{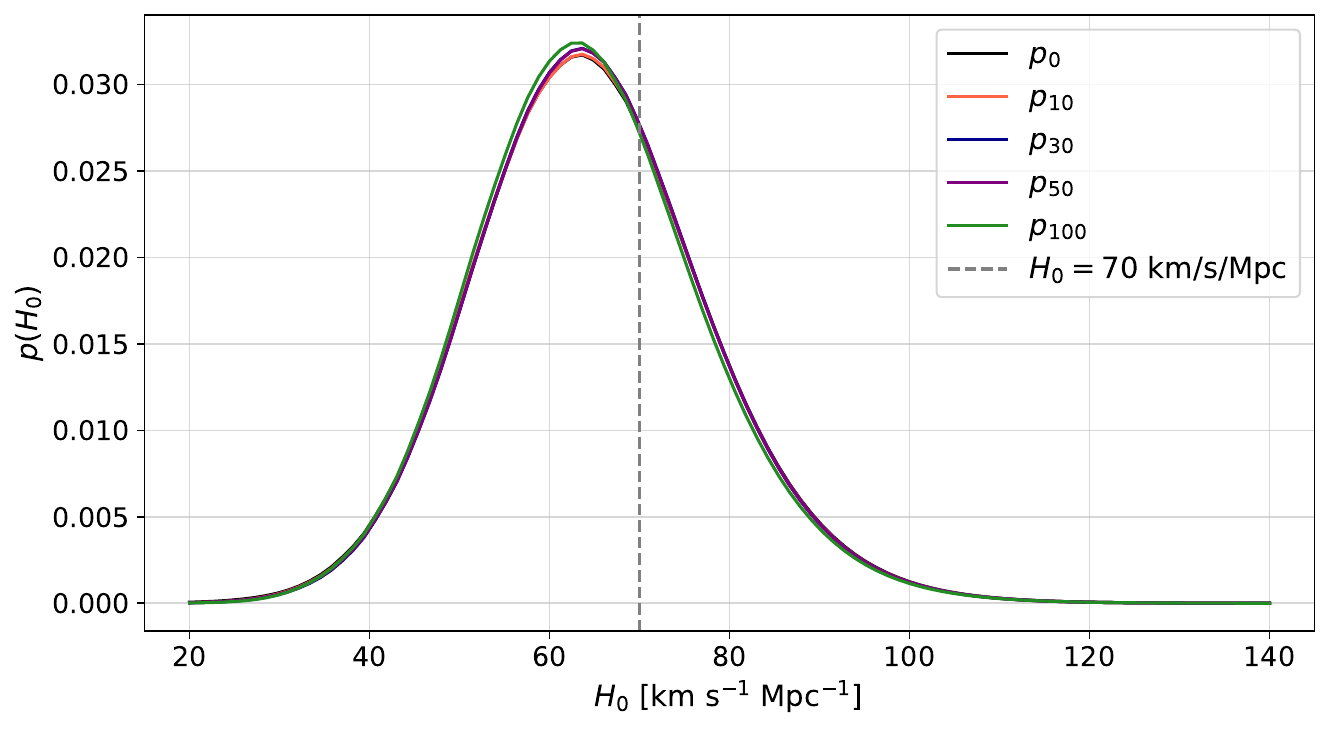}
  \caption{Joint $H_0$ posterior from 42 BBHs for completeness percentiles 10\% (red), 30\% (blue), 50\% (darkblue), and 100\% (green). The vertical dashed line marks $H_0 = 70~\mathrm{km~s^{-1}~Mpc^{-1}}$.}
  \label{fig:Multievent_nside32}
\end{figure}

The results of the joint analysis of GWTC-3 42 BBHs are shown in Fig.~\ref{fig:Multievent_nside32} and Table ~\ref{Tab:Multievent_nside32}.

\begin{figure}[h]
  \centering
  \includegraphics[width=0.85\textwidth]{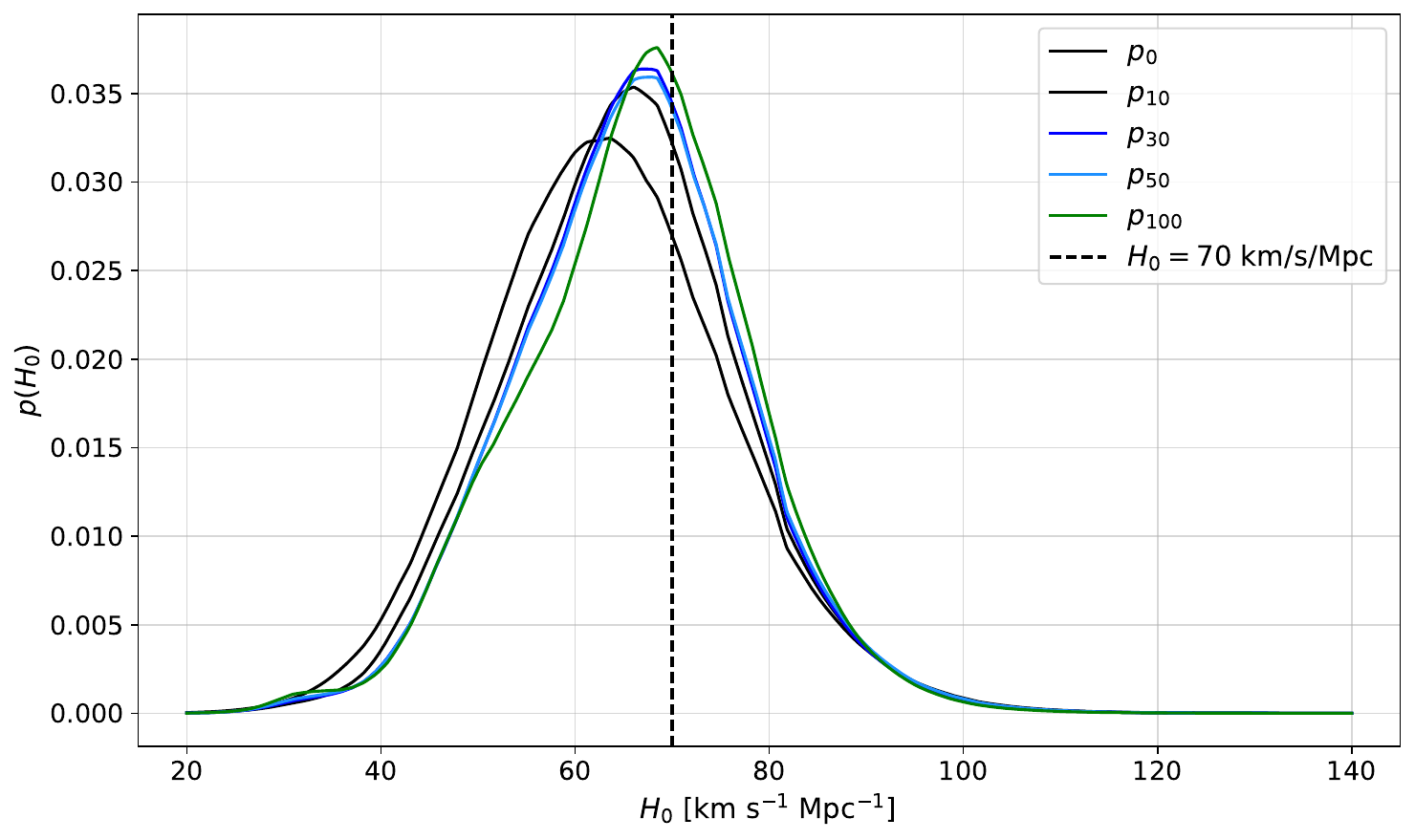}
  \caption{Combined $H_0$ posterior for 46 events (42 BBHs + 3 NSBHs + GW190425) for different value of the completeness percentile. Posteriors are obtained by multiplying the single-event posteriors computed with LOS redshift priors at each percentile. The vertical dashed line marks $H_0 = 70~\mathrm{km~s^{-1}~Mpc^{-1}}$.}
  \label{fig:Combining_nside32}
\end{figure}

\begin{table*}[h]
\centering
\caption{Best--fit values and 68\% and 95\% confidence intervals for the combined $H_0$ posterior distribution using different completeness percentiles of the catalog. We report the best--fit estimate, the 68\% and 95\% credible intervals, and the relative errors, defined as the width of the interval divided by the best--fit value.}
\begin{tabular}{lccccc}
\hline
Percentile & $\boldsymbol{H_0}$ [km s$^{-1}$ Mpc$^{-1}$] & 68\% CI & 95\% CI & Rel.\ Err.\ (68\%) & Rel.\ Err.\ (95\%) \\
\hline
0   & $63.64^{+13.20}_{-12.24}$ & (51.39, 76.85) & (39.80, 91.83) & 39.98\% & 81.77\% \\
10  & $63.64^{+13.14}_{-12.24}$ & (51.40, 76.78) & (40.03, 91.68) & 39.88\% & 81.17\% \\
30  & $63.64^{+13.12}_{-12.58}$ & (51.06, 76.76) & (40.32, 91.61) & 39.54\% & 80.61\% \\
50  & $63.64^{+13.10}_{-12.08}$ & (51.56, 76.74) & (40.21, 91.65) & 39.60\% & 80.85\% \\
100 & $63.64^{+12.60}_{-12.37}$ & (51.27, 76.24) & (40.16, 91.21) & 39.24\% & 80.22\% \\
\hline
\label{Tab:Multievent_nside32}
\end{tabular}
\end{table*}
In Fig.~\ref{fig:Combining_nside32} and Table~\ref{Tab:Combining_nside32} we show the results of the combined single event posterior of all GWs events.

\begin{table*}[h]
\centering
\caption{Best--fit estimates of the combined $H_0$ posterior using different catalog--completeness percentiles (46 events). We report the best--fit value with 68\% errors, the central 68\% and 95\% credible intervals, and the relative errors (interval width divided by the best--fit value).}
\begin{tabular}{lccccc}
\hline
Percentile & $\boldsymbol{H_0}$ [km s$^{-1}$ Mpc$^{-1}$] & 68\% CI & 95\% CI & Rel.\ Err.\ (68\%) & Rel.\ Err.\ (95\%) \\
\hline
0   & $63.60^{+12.08}_{-12.56}$ & (51.04, 75.68) & (39.77, 89.93) & 38.74\% & 78.86\% \\
10  & $66.01^{+10.48}_{-12.97}$ & (53.04, 76.49) & (42.02, 89.79) & 35.53\% & 72.37\% \\
30  & $67.09^{+9.75}_{-13.06}$  & (54.03, 76.84) & (42.84, 89.51) & 34.01\% & 69.57\% \\
50  & $67.69^{+9.42}_{-13.70}$  & (53.99, 77.11) & (42.71, 89.93) & 34.15\% & 69.77\% \\
100 & $68.41^{+9.10}_{-14.15}$  & (54.26, 77.51) & (42.49, 89.16) & 33.99\% & 68.22\% \\
\hline
\label{Tab:Combining_nside32}
\end{tabular}
\end{table*}

\begin{figure}[t]
    \centering
    \includegraphics[width=0.7\textwidth]{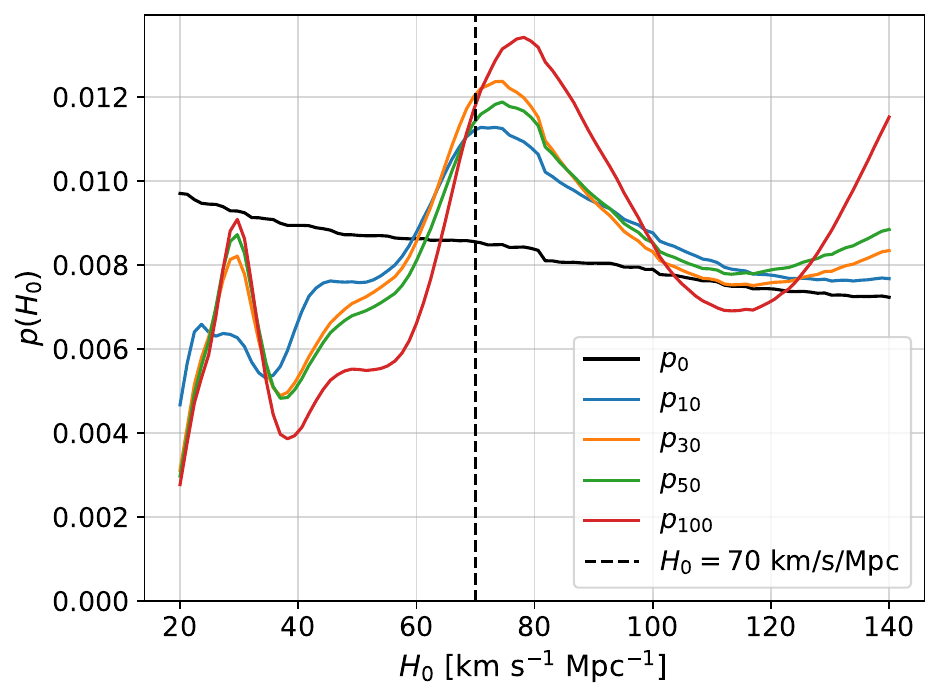}
    \caption{Posterior distributions of $H_{0}$ for GW190814 using LOS priors computed at 
    different completeness percentiles. The vertical dashed line indicates $H_{0}=70~\mathrm{km\,s^{-1}\,Mpc^{-1}}$.}
    \label{fig:GW190814_nside32}
\end{figure}

In Fig.~\ref{fig:GW190814_nside32} and Table~\ref{Tab:GW190814_nside32} we show the posteriors for the  event GW190814.

\begin{table*}[htbp]
    \centering
    \caption{Best--fit values of $H_{0}$ for GW190814 at different completeness percentiles, 
    with $68\%$ and $95\%$ central credible intervals and their relative errors.}
    \begin{tabular}{lccccc}
        \hline
        Percentile 
        & $\boldsymbol{H_0}$ [km s$^{-1}$ Mpc$^{-1}$] 
        & 68\% CI 
        & 95\% CI 
        & Rel.\ Err.\ (68\%) 
        & Rel.\ Err.\ (95\%) \\
        \hline
        0   
        & $20.00^{+98.13}_{-15.91}$ 
        & $(35.91,\ 118.13)$ 
        & $(21.37,\ 136.55)$ 
        & $411.14\%$ 
        & $575.90\%$ \\

        10  
        & $73.33^{+45.80}_{-28.77}$ 
        & $(44.56,\ 119.13)$ 
        & $(23.14,\ 136.74)$ 
        & $101.69\%$ 
        & $154.91\%$ \\

        30  
        & $74.55^{+45.23}_{-29.17}$ 
        & $(45.38,\ 119.78)$ 
        & $(24.10,\ 136.99)$ 
        & $99.80\%$ 
        & $151.43\%$ \\

        50  
        & $74.55^{+46.27}_{-29.23}$ 
        & $(45.32,\ 120.82)$ 
        & $(24.26,\ 137.16)$ 
        & $101.28\%$ 
        & $151.45\%$ \\

        100 
        & $78.18^{+44.79}_{-30.64}$ 
        & $(47.54,\ 122.97)$ 
        & $(24.47,\ 137.80)$ 
        & $96.48\%$ 
        & $144.96\%$ \\
        \hline
    \end{tabular}
    \label{Tab:GW190814_nside32}
\end{table*}

\begin{figure*}[t]
  \centering
  \includegraphics[width=\textwidth]{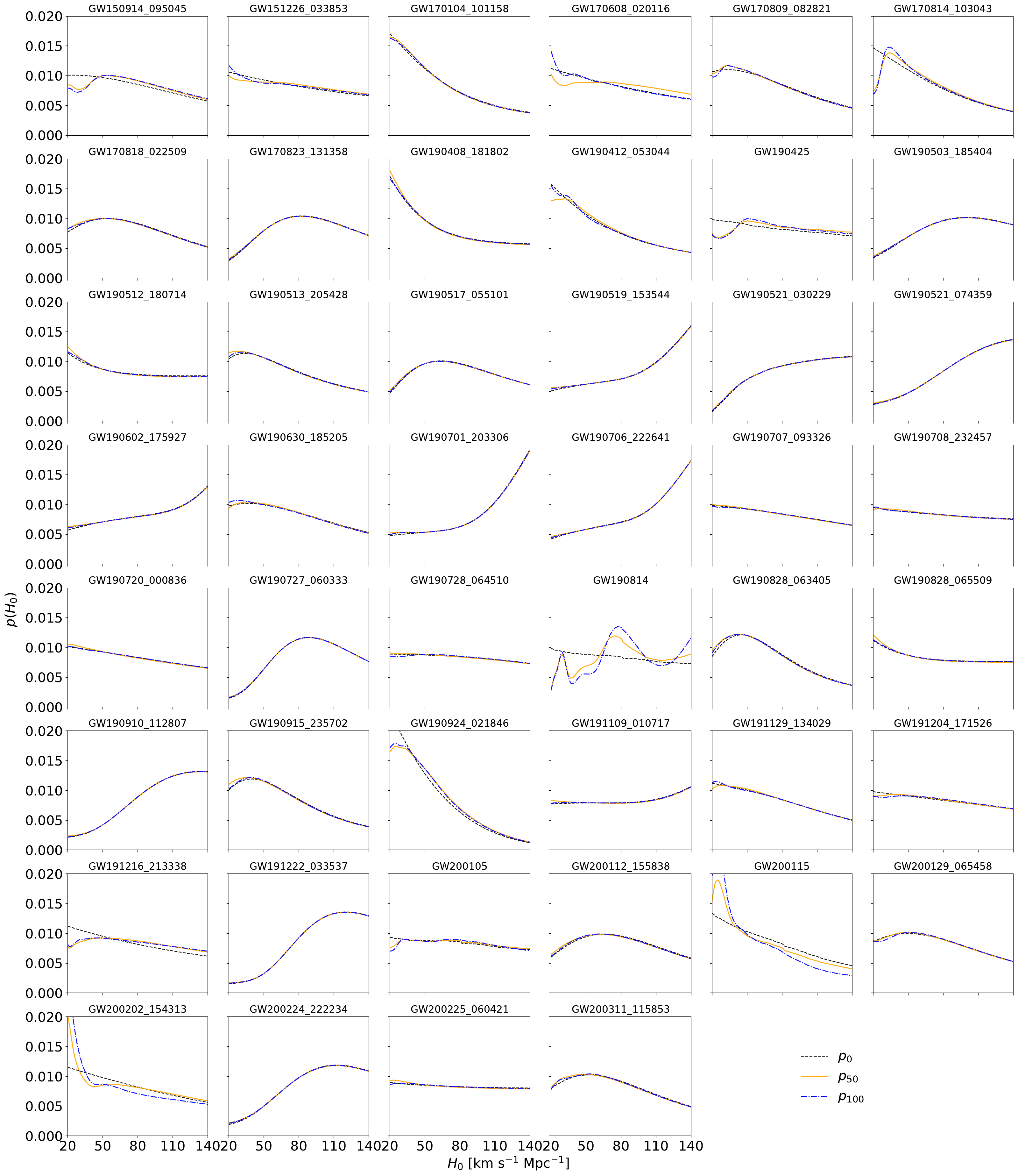}
  \caption{Individual $H_0$ posteriors for 42 BBHs,  3 NSBHs and GW190425  for different LOS obtained different values of the completeness percentile. Each panel shows one event; curves of different colors correspond to different percentiles.}
  \label{Fig:Individual_nside32}
\end{figure*}

The Fig. \ref{Fig:Individual_nside32} show the $H_0$ posterior for each event.

\section{Measures of the difference between $p_0(z)$ and $p_{t}(z)$ for individual events}
\label{sec:Indicators_appendix}
The values of different measures of the difference between $p_0(z)$ and and $p_t(z)$ for different values of the completeness $t$ is reported in Table~\ref{tab:indicators_50} and Table~\ref{tab:indicators_100}.

\begin{table*}[t]
\centering
\caption{Measures of the difference between $p_0(z)$ and $p_{50}(z)$ for nside=64}
\label{tab:indicators_50}
\begin{tabular}{lrrrrrr}
\toprule
Event & $\Omega_{90}$ (deg$^{2}$) & KL & Hellinger \\
\midrule
GW150914 & 251.7488 & 0.003973 & 0.032681 \\
GW151226 & 950.3427 & 0.002169 & 0.023701 \\
GW170104 & 1012.5684 & 0.001891 & 0.022129 \\
GW170608 & 378.6526 & 0.001664 & 0.020786 \\
GW170809 & 275.6817 & 0.002324 & 0.024716 \\
GW170814 & 92.1420 & 0.001137 & 0.017024 \\
GW170818 & 32.4341 & 0.002550 & 0.026148 \\
GW170823 & 1783.6695 & 0.001963 & 0.022574 \\
GW190408 & 293.5102 & 0.002958 & 0.027771 \\
GW190412 & 152.3220 & 0.002668 & 0.026714 \\
GW190503 & 100.3907 & 0.002081 & 0.023236 \\
GW190512 & 286.6844 & 0.003482 & 0.030363 \\
GW190513 & 445.7830 & 0.002063 & 0.023105 \\
GW190517 & 506.6776 & 0.001622 & 0.020453 \\
GW190519 & 568.0182 & 0.002186 & 0.023801 \\
GW190521\_030229 & 1020.5155 & 0.001801 & 0.021623 \\
GW190521\_074359 & 466.9227 & 0.002119 & 0.023567 \\
GW190602 & 761.6065 & 0.001418 & 0.019157 \\
GW190630 & 956.0210 & 0.001951 & 0.022530 \\
GW190701 & 44.6498 & 0.002544 & 0.025717 \\
GW190706 & 2924.4138 & 0.001767 & 0.021360 \\
GW190707 & 1156.8089 & 0.001798 & 0.021565 \\
GW190708 & 11514.0597 & 0.001607 & 0.020367 \\
GW190720 & 256.1977 & 0.002514 & 0.025906 \\
GW190727 & 381.4917 & 0.003195 & 0.029026 \\
GW190728 & 399.7464 & 0.001753 & 0.021373 \\
GW190814 & 18.4940 & 0.005002 & 0.036411 \\
GW190828\_063405 & 339.8254 & 0.002498 & 0.025723 \\
GW190828\_065509 & 593.2364 & 0.002143 & 0.023654 \\
GW190910 & 9647.8903 & 0.001727 & 0.021140 \\
GW190915 & 449.2188 & 0.001960 & 0.022585 \\
GW190924 & 377.4592 & 0.002578 & 0.026064 \\
GW191109 & 1784.4432 & 0.001689 & 0.020969 \\
GW191129 & 848.4734 & 0.001762 & 0.021332 \\
GW191204 & 345.1333 & 0.002106 & 0.023332 \\
GW191216 & 480.3382 & 0.002256 & 0.024234 \\
GW191222 & 2299.4679 & 0.001714 & 0.021131 \\
GW200105 & 7882.0164 & 0.001752 & 0.021310 \\
GW200112 & 4250.6027 & 0.001952 & 0.022515 \\
GW200115 & 512.2052 & 0.002363 & 0.024770 \\
GW200129 & 80.5033 & 0.002325 & 0.024720 \\
GW200202 & 159.2232 & 0.002932 & 0.028055 \\
GW200224 & 49.6593 & 0.001857 & 0.021805 \\
GW200225 & 509.2021 & 0.001659 & 0.020673 \\
GW200311 & 35.1913 & 0.002390 & 0.024989 \\
GW200316 & 410.5392 & 0.002887 & 0.027616 \\
\bottomrule
\end{tabular}
\end{table*}
\nocite{*}

\begin{table*}[t]
\centering
\caption{Measures of the difference between $p_0(z)$ and $p_{100}(z)$ for nside=64}
\label{tab:indicators_100}
\begin{tabular}{lrrrrrr}
\toprule
   Event &   $\Omega_{90}$ (deg$^{2}$) &     KL &   Hellinger \\
\midrule
GW150914 & 251.7488 & 0.004460 & 0.034203 \\
GW151226 & 950.3427 & 0.001354 & 0.018376 \\
GW170104 & 1012.5684 & 0.000939 & 0.015343 \\
GW170608 & 378.6526 & 0.000838 & 0.014345 \\
GW170809 & 275.6817 & 0.002269 & 0.024242 \\
GW170814 & 92.1420 & 0.001214 & 0.017420 \\
GW170818 & 32.4341 & 0.002028 & 0.023149 \\
GW170823 & 1783.6695 & 0.001077 & 0.016543 \\
GW190408 & 293.5102 & 0.001561 & 0.019699 \\
GW190412 & 152.3220 & 0.003873 & 0.030914 \\
GW190503 & 100.3907 & 0.002450 & 0.024915 \\
GW190512 & 286.6844 & 0.003399 & 0.029789 \\
GW190513 & 445.7830 & 0.000985 & 0.015773 \\
GW190517 & 506.6776 & 0.001236 & 0.017579 \\
GW190519 & 568.0182 & 0.001038 & 0.016108 \\
GW190521\_030229 & 1020.5155 & 0.001023 & 0.016080 \\
GW190521\_074359 & 466.9227 & 0.001037 & 0.016303 \\
GW190602 & 761.6065 & 0.001135 & 0.017025 \\
GW190630 & 956.0210 & 0.001213 & 0.017506 \\
GW190701 & 44.6498 & 0.003217 & 0.027846 \\
GW190706 & 2924.4138 & 0.000737 & 0.013613 \\
GW190707 & 1156.8089 & 0.001013 & 0.015961 \\
GW190708 & 11514.0597 & 0.000689 & 0.013124 \\
GW190720 & 256.1977 & 0.003438 & 0.029137 \\
GW190727 & 381.4917 & 0.002121 & 0.023368 \\
GW190728 & 399.7464 & 0.000781 & 0.014059 \\
GW190814 & 18.4940 & 0.005309 & 0.037188 \\
GW190828\_063405 & 339.8254 & 0.001806 & 0.021680 \\
GW190828\_065509 & 593.2364 & 0.001898 & 0.022112 \\
GW190910 & 9647.8903 & 0.000904 & 0.015145 \\
GW190915 & 449.2188 & 0.001382 & 0.018440 \\
GW190924 & 377.4592 & 0.001678 & 0.020778 \\
GW191109 & 1784.4432 & 0.001300 & 0.018177 \\
GW191129 & 848.4734 & 0.001073 & 0.016424 \\
GW191204 & 345.1333 & 0.002046 & 0.022710 \\
GW191216 & 480.3382 & 0.001031 & 0.016135 \\
GW191222 & 2299.4679 & 0.001018 & 0.016083 \\
GW200105 & 7882.0164 & 0.000812 & 0.014352 \\
GW200112 & 4250.6027 & 0.000962 & 0.015612 \\
GW200115 & 512.2052 & 0.001725 & 0.020893 \\
GW200129 & 80.5033 & 0.002047 & 0.023011 \\
GW200202 & 159.2232 & 0.002453 & 0.025459 \\
GW200224 & 49.6593 & 0.002917 & 0.026085 \\
GW200225 & 509.2021 & 0.000896 & 0.014920 \\
GW200311 & 35.1913 & 0.002244 & 0.023420 \\
GW200316 & 410.5392 & 0.002278 & 0.023682 \\
\bottomrule
\end{tabular}
\end{table*}

\end{document}